\documentclass[12pt]{article}
\usepackage{graphicx}
\usepackage{amsmath, amssymb, amsfonts} 
\usepackage{bm}
\usepackage{accents}
\usepackage{subfig}
\usepackage{enumitem}
\usepackage{natbib} 

\newcommand{\ba}{\boldsymbol a}
\newcommand{\bb}{\boldsymbol b}
\newcommand{\bc}{\boldsymbol c}

\newcommand{\br}{\boldsymbol r}
\newcommand{\bx}{\boldsymbol x}
\newcommand{\bA}{\boldsymbol A}
\newcommand{\bB}{\boldsymbol B}
\newcommand{\bC}{\boldsymbol C}
\newcommand{\bE}{\boldsymbol E}

\newcommand{\bX}{\boldsymbol X}

\newcommand{\bSigma}{\boldsymbol \Sigma}

\newcommand{\bha}{\bm{{\hat{a}}}}
\newcommand{\bhA}{\bm{{\hat{A}}}}
\newcommand{\bhB}{\bm{{\hat{B}}}}
\newcommand{\bhC}{\bm{{\hat{C}}}}
\newcommand{\bhX}{\bm{{\hat{X}}}}
\newcommand{\bhSigma}{\bm{{\hat{\Sigma}}}}

\newcommand{\bcx}{\accentset{\circ}{\bx}}
\newcommand{\bcX}{\accentset{\circ}{\bX}}

\newcommand{\sbullet}{\hbox{\fontfamily{lmr}\fontsize{10}{0}\selectfont\textbullet}}
\newcommand{\bfx}{\accentset{\sbullet}{\bx}}
\newcommand{\bfX}{\accentset{\sbullet}{\bX}}

\newcommand{\btE}{\bm{{\Tilde{E}}}}

\newcommand{\btx}{\bm{{\Tilde{x}}}}
\newcommand{\btX}{\bm{{\Tilde{X}}}}

\newcommand{\ubE}{\underline{\boldsymbol{E}}}
\newcommand{\ubR}{\underline{\boldsymbol{R}}}
\newcommand{\ubX}{\underline{\boldsymbol{X}}}
\newcommand{\ubcX}{\accentset{\circ}{\underline{\bX}}}
\newcommand{\ubhX}{\underline{\bhX}}
\newcommand{\ubfX}{\accentset{\sbullet}{\underline{\bX}}}

\newcommand{\ubtX}{\underline{\btX}}

\newcommand{\bfrd}{\accentset{\sbullet}{\text{\footnotesize RD}}}	
\def\mcd{\text{\tiny MCD}}

\def\rd{\text{\tiny RD}}
\def\RD{\text{RD}}
\newcommand{\tRD}{\accentset{\sim}{\RD}}
\def\sd{\text{\tiny SD}}
\def\SD{\text{SD}}
\def\DDC{\text{\tiny DDC}}
\newcommand{\R}{\mathbb{R}}

\setlist{
    topsep=3pt,
    itemsep=1pt,
    parsep=1pt
}
\begin{document}

\def\spacingset#1{\renewcommand{\baselinestretch}%
{#1}\small\normalsize} \spacingset{1}


\title{\bf MacroPARAFAC for handling rowwise and cellwise outliers in incomplete multi-way data}
\author{Mia Hubert and Mehdi Hirari\\
	      \\
    Section of Statistics and Data Science, \\
    Department of Mathematics, KU Leuven
    }

\date{July 5, 2024}

\maketitle

\begin{abstract}
Multiway data extend two-way matrices into higher-dimensional tensors, often explored through dimensional reduction techniques. In this paper, we study the Parallel Factor Analysis (PARAFAC) model for handling multiway data, representing it more compactly through a concise set of loading matrices and scores. We assume that the data may be incomplete and could contain both rowwise and cellwise outliers, signifying cases that deviate from the majority and outlying cells dispersed throughout the data array. To address these challenges, we present a novel algorithm designed to robustly estimate both loadings and scores. Additionally, we introduce an enhanced outlier map to distinguish various patterns of outlying behavior. Through simulations and the analysis of fluorescence Excitation-Emission Matrix (EEM) data, we demonstrate the robustness of our approach. Our results underscore the effectiveness of diagnostic tools in identifying and interpreting unusual patterns within the data.\\
\end{abstract}

\noindent%

\bigskip

\spacingset{1.2} 

\section{Introduction}
\label{sec:intro}

PARAllel FACtor (PARAFAC) analysis can be seen as an extension of principal component analysis to multiway or tensor data. It decomposes a data array into its building components. Originally the methodology and applications were developed in psychometrics \citep{Kroonenberg:BookMulti}, but nowadays the method is also very  popular in engineering and chemometrics \citep{Smilde:Bookmultiway}. 
Applications can for example be found in structural damage detection, traffic data imputation and flow prediction, signal processing and computer vision \citep{Goldfarb:RobustTensor, Salehi:ReviewTensor, Schmitz:PARAFAC}. In chemometrics, PARAFAC is often used to analyse fluorescence excitation-emission matrices (EEM) as it is able to decompose a mixture of fluorophores into its chemical components \citep{Bro:parafac, Murphy:PARAFAC}. 

In this paper we focus on three-way data although the methodology could also be extended to higher order tensors. We assume that the given data array $\underline{\boldsymbol{X}}$ has dimensions $I \times J \times K$, where $i=1,\dots,I$ runs over the observations. The measurements of the $i$th observation are thus contained in a matrix $\boldsymbol{X}_{i}$ of size $J \times K$. 

Whereas most work on PARAFAC assumes that the data are homogeneous and without contamination, we make some more challenging assumptions. First we do not require that the data array is fully observed, i.e.\ it may contain missing values (NAs) according to MAR (Missing at Random). 

Next we allow that some observations are outlying, i.e.\ some of the slices $\bX_i$ may belong to another population than the majority of the samples. This type of outliers is often studied in the robust statistics literature. A general name for it is \textit{casewise} contamination, which we could adapt here as well. For consistency with earlier work, we call them \textit{rowwise outliers}, which originates from the fact that many statistical methods operate on multivariate data, where observations align with rows in the data matrix. Here, observations correspond with rows of the unfolded array $\boldsymbol{X}^{I \times JK}$, where each row $\bx'_i$ of length $JK$ is the vectorized slice $\bX_i$.

Finally we allow individual values $x_{ijk}$ (cells in the data array) to be outlying. These so-called \textit{cellwise outliers} may occur in some of the slices $\bX_i$ or even in all of them, hence contaminating all the observations. Each slice can have none, a few or many outlying cells. This problem has only been studied more recently and is much harder to tackle. Rowwise robust methods typically try to detect the rowwise outliers and eliminate or downweight them during the estimation procedure. When a row contains only a few outlying values and is designated as a rowwise outlier, valuable information embedded in the uncontaminated values risks to be discarded, which leads to a loss of efficiency. In multiway data this effect is even more pronounced as the number of cells $JK$ for each case might be very large, hence it is strongly preferred to remove or downweight the outlying cells only instead of downweighting the whole slice. 

As far as we know, currently no methods are available that can cope with the three issues (NAs, rowwise and cellwise outliers) simultaneously. Algorithms are available for incomplete data \citep{Tomasi:PARAFACmissing} but they are not robust to any type of contamination. A rowwise robust method (RPARAFAC) has been proposed in  \cite{Engelen:robParafac}. It has been algoritmically adapted in \cite{Todorov:PARAFAC} and extended to compositional data in \cite{DiPalma:PARAFACcompositional} but it can not handle missing values nor cellwise outliers. The improved RPARAFAC-SI method \citep{Hubert:RParafac-SI} allows for missing values as well but still not for cellwise contamination. On the other hand in \cite{Goldfarb:RobustTensor} and \cite{Heng:RobustTensor} algorithms have been constructed to deal with cellwise outliers and NAs but their methods cannot resist rowwise contamination. 

In this paper we present MacroPARAFAC where `Macro' stands for \textbf{M}issingness \textbf{a}nd \textbf{c}ellwise and \textbf{r}owwise \textbf{o}utliers, similar to the acronym MacroPCA for robust PCA \citep{Hubert:MacroPCA}. 
Section~\ref{sect:PARAFAC} shortly reviews classical PARAFAC and its alternating least-squares algorithm. Our new algorithm is proposed in Section~\ref{sect:MacroPARAFAC} followed by a simulation study in Section~\ref{sect:simul}. Outlier detection tools, including a new enhanced outlier map, are described in Section~\ref{sect:outlierdetection} and applied to a real data set in Section~\ref{sect:realdata}.  

\section{Classical PARAFAC} 
\label{sect:PARAFAC}
The PARAFAC model for three-way data $\ubX \in \R^{I \times J \times K}$ assumes that the data array $\ubX$ can be well approximated by a sum of $F$ rank-1 tensors. If we define the $(I \times F)$ score matrix $\bA$ with columns $\ba_1, \ldots,\ba_F$ and the two loading matrices $\boldsymbol{B} = (\bb_1 \, \dots \, \bb_F)$ and $\boldsymbol{C}=(\bc_1 \, \dots \, \bc_F)$ of size $(J \times F)$ and $(K \times F)$ respectively, then the model states that
\begin{equation} \label{eq:model}
\ubX = \sum_{f=1}^F \ba_f \otimes \bb_f \otimes \bc_f + \ubE
\end{equation}
with $\ubE$ containing the errors and $\otimes$ denoting the outer product between vectors. Unfolding or matricizing the data array along the first mode yields a matrix $\bX = \boldsymbol{X}^{I \times JK}$. Then model~\eqref{eq:model} can also be written as
\begin{equation*}
\boldsymbol{X}^{I \times J K}=\boldsymbol{A}(\boldsymbol{C} \odot \boldsymbol{B})^{\prime}+\boldsymbol{E}^{I \times J K},
\end{equation*}
with $(\boldsymbol{C} \odot \boldsymbol{B}) = [vec(\bb_1 \bc_1') \, \dots \, vec(\bb_F \bc_F')]$
the Khatri-Rao product between the loading matrices and 
$\boldsymbol{E}^{I \times J K}$ the unfolded error matrix. 

Classical PARAFAC estimates the scores and loadings by minimizing the loss function:
\begin{align*}
\|\boldsymbol{X}-\bhX\|_{F}^{2} &=\|\boldsymbol{X}-\bhA(\bhC \odot \bhB)^{\prime}\|_{F}^{2} \\
&=\sum_{i=1}^{I} \sum_{j=1}^{J} \sum_{k=1}^{K}\left(x_{i j k}-\hat{x}_{i j k}\right)^{2} \,.
\end{align*}
If we define the residual distance (RD) of observation $i$ as
\begin{equation*}
    \RD_i = \|\bX_i - \bhX_i \|_F = \sqrt{\sum_{j=1}^J \sum_{k=1}^k (x_{ijk}-\hat{x}_{ijk})^2}
\end{equation*}
then it corresponds with minimizing the sum of the squared residual distances.
Typically this minimization is performed using an alternating least squares algorithm PARAFAC-ALS which is implemented in the R package \texttt{multiway} \citep{R, R:multiway} that we used to perform our computations, see e.g.\ \cite{Bro:parafac}. In this implementation initial estimates for the loading matrices can be provided. If not, a random initialisation is used. 
It is also possible to add constraints to the estimates, such as non-negativity or smoothness of the loadings.

An overview of the standard PARAFAC-ALS algorithm is as follows. Given the array $\ubX$, initial loadings $\boldsymbol{B}^{(0)}$ and $\boldsymbol{C}^{(0)}$, for $s \geqslant 1$: 

\begin{enumerate}
    \item
     Estimate $\boldsymbol{A}^{(s)}$ from $\ubX, \boldsymbol{B}^{(s-1)}$ and $\boldsymbol{C}^{(s-1)}$ by least squares regression
    \item 
     Estimate $\boldsymbol{B}^{(s)}$ from $\ubX, \boldsymbol{A}^{(s)}$ and $\boldsymbol{C}^{(s-1)}$ similarly
    \item 
     Estimate $\boldsymbol{C}^{(s)}$ from $\ubX, \boldsymbol{A}^{(s)}$ and $\boldsymbol{B}^{(s)}$ similarly
  \end{enumerate}
This iteration is repeated until convergence.

When $\ubX$ contains missing values, this algorithm can be easily adapted by first imputing the missing values (e.g.\ with the average values of the observed values in each column of $\bX$) and then updating them after each iteration \citep{Tomasi:PARAFACmissing}. More precisely, two steps are then added to the ALS algorithm:
\begin{enumerate}
    \item[4.] Compute $\bhX^{(s)}=\bhA^{(s)}(\bhC^{(s)} \odot \bhB^{(s)})^{\prime}$
    \item[5.] Replace the NAs from $\ubX$ by the corresponding cells from $\ubhX^{(s)}$\,.
\end{enumerate}
Steps 1-5 are then iterated until convergence. 

\section{The MacroPARAFAC algorithm} 
\label{sect:MacroPARAFAC}
To construct MacroPARAFAC we closely follow the notations and the principles underlying MacroPCA for multivariate data.  
It is assumed that at most $I-h$ slices (with $\lceil I/2 \rceil  < h < I$) 
are outlying cases, i.e.\ observations that deviate from the  majority. The other $h$ slices can have outlying cells, but imputing these cells with appropriate values transforms them into well-fitting observations. 
The set of potentially cellwise outliers is fixed after the first stage of the algorithm but their imputed values are updated several times. On the other hand the set of potentially rowwise outliers can change throughout the algorithm, and these observations are always excluded to estimate the loading matrices.

We use the following notations to distinguish the various types of imputations applied in the course of the algorithm:
\begin{itemize}
    \item When in an array, matrix, or vector only missing values are imputed, we denote this with an open circle as  superscript. Consequently, the \textit{NA-imputed} data array is denoted as $\ubcX$, the unfolded NA-imputed matrix as $\bcX$, and so forth. 
    \item The \textit{cell-imputed} array is denoted as $\ubfX$. It has imputed values for all missing values, and  each matrix $\bfX_{i}$ has imputed values for the outlying cells that do not belong to an outlying row.
    \item The \textit{fully imputed} array $\ubtX$ has imputed values for all outlying cells and all missing values.
\end{itemize}

MacroPARAFAC comprises the following stages:

\begin{enumerate}
    \item Initialization of NAs, cellwise outliers and rowwise outliers: 
To start, we first apply the DetectDeviatingCells (DDC) algorithm of \cite{Rousseeuw:DDC} to the unfolded matrix $\boldsymbol{X}^{I \times J K}$. Since in general $JK$ can be quite large, we adopt the FastDDC algorithm constructed by \cite{Raymaekers:FastCorr} from the \texttt{cellWise} R package \citep{R:cellWise}.
DDC provides imputations for all missing values, hence the initial NA-imputed matrix $\bcX^{(0)}$ is obtained by replacing the NAs in $\bX$ by their DDC estimates.  
DDC also yields the positions of cellwise outliers $I_{c, \text{\tiny DDC}}$ and imputed values for these cells, from which we can compute the initial fully imputed matrix $\btX^{(0)}$. Finally DDC provides indices $I_{r, \text{\tiny DDC}}$ of potential outlying rows, albeit not necessarily flagging all of them. We then construct the initial cell-imputed data matrix $\bfX^{(0)}$ by imputing the missing values in all rows and the cellwise outliers in the $h$ rows with the fewest cells flagged by DDC but not in $I_{r, \text{\tiny DDC}}$. The $I-h$ remaining rows constitute the first potential set of rowwise outliers $I_{r,\text{par}}$. 

\item Improving the potential set of rowwise outliers: Next, we compute the multivariate outlyingness of a case $\bfx_i^{(0)}$, the $i$th row of $\bfX^{(0)}$, as: 
\begin{equation}  \label{eq:outl}
\text { outl}\left(\bfx_i^{(0)}\right)=\max _{\boldsymbol{v} \in D} \frac{\left|\boldsymbol{v}^{\prime} \bfx_i^{(0)}-\hat{\mu}_{\mcd}\left(\boldsymbol{v}^{\prime} \bfx_j^{(0)}\right)\right|}{\hat{\sigma}_{\mcd}\left(\boldsymbol{v}^{\prime} \bfx_j^{(0)}\right)}
\end{equation}
where $\hat{\mu}_{\mcd}\left(\boldsymbol{v}^{\prime} \bfx_j^{(0)}\right)$ and $\hat{\sigma}_{\mcd}\left(\boldsymbol{v}^{\prime} \bfx_j^{(0)}\right)$ are the univariate Minimum Covariance Determinant (MCD) location and scale estimates of the $(\boldsymbol{v}^{\prime} \bfx_1^{(0)}, \dots, \boldsymbol{v}^{\prime} \bfx_I^{(0)}$) \citep{Hubert:WIRE-MCD2}, and the set $D$ contains 250 directions $\boldsymbol{v}$ through two vectors $\bfx_i^{(0)}$ \citep{Hubert:ROBPCA}. The index set of the $h$ cases with smallest outlyingness (but not in $I_{r,\DDC}$) is denoted as $H_0$. The complementary set serves as the new $I_{r,\text{\tiny par}}$.
\item
Initialisation of the PARAFAC loadings: A new cell-imputed matrix $\bfX^{(1)}$ is constructed by imputing the NAs in all rows and the outlying cells in the rows of $H_0$ with the corresponding values from $\btX^{(0)}$. This means that $\bfx_i^{(1)} = \btx_i^{(0)}$ for 
$i \in H_0$\,, and $\bfx_i^{(1)} = \bcx_i^{(0)}$ if 
$i \notin H_0$. We then apply PARAFAC-ALS for complete data to $\bfX^{(1)}_{H_0}$, the $h$ slices $\bfX^{(1)}_{i}$ with $i \in H_{0}$, using random starts for the loading matrices. 
This allows to determine the number of components needed to pursue the analysis, for example using the core consistency diagnostic or via a split-half analysis \citep{Murphy:PARAFAC}.
Further we obtain initial loading matrices $\bhB^{(1)}$ and $\bhC^{(1)}$. We also obtain scores $\bhA^{(1)}_{H_0} = \bfX^{(1)}_{H_0} (\bhC^{(1)} \odot \bhB^{(1)})^{\dagger}$ (where $\dagger$ denotes the Moore-Penrose inverse)
and fitted values $\bhX_{H_0} = \bhA^{(1)}_{H_0}(\bhC^{(1)} \odot \bhB^{(1)})^{\prime}$.
We then update $\bfX^{(1)}_{H_0}$ by replacing the cells corresponding to NAs and cellwise outliers with the current fitted values. 

\item
Iterative estimation: The goal of this step is to estimate the loadings and scores iteratively while improving on the imputations. We apply PARAFAC-ALS for incomplete data to $\bfX^{(1)}_{H_0}$ starting with $\bhB^{(1)}$ and $\bhC^{(1)}$. In the last step of each iteration, both the NAs and cellwise outliers are updated. At convergence, we have loadings $\bhB^{(2)}$ and $\bhC^{(2)}$, as well as scores $\bha_i^{(2)}$ for the observations $i \in H_0$. We then also compute scores for the remaining cases $i \notin H_0$ as outlined in \cite{Smilde:Bookmultiway}:
\begin{equation*}
 \bha_i^{(2)} = (\bhC^{(2)} \odot \bhB^{(2)})^{\dagger} \btx^{(0)}_i \,.
\end{equation*}
The fitted values 
$\bhX = \bhA^{(2)}(\bhC^{(2)} \odot \bhB^{(2)})^{\prime}$
are then again used to update the NAs and cellwise outliers of all samples, yielding $\bcX^{(2)}, \bfX^{(2)}$ and $\btX^{(2)}$.

\item
Re-weighting: The re-weighting step aims to include more observations in the estimation step to improve the efficiency at a low computational cost. 
First we recompute the scores and fitted values 
of the cell-imputed data as $\bhA^{(3)} = \bfX^{(2)} (\bhC^{(2)} \odot \bhB^{(2)})^{\dagger}$ and $\bhX = \bhA^{(3)}(\bhC^{(2)} \odot \bhB^{(2)})^{\prime}$.
Then we compute for all observations their residual distance:
\begin{equation*}
\bfrd_i=\left\|\bfX_{i}^{(2)}-\bhX_{i}\right\|_{F} \,.
\end{equation*}
For regular observations we can assume that their residual distances to the power $2/3$ are roughly Gaussian \citep{Engelen:robParafac}. Hence we derive the cutoff value:
\begin{equation} \label{eq:cutoff_rd}
c_{\rd}:=\left(\hat{\mu}_{\mcd}\left(\left\{\bfrd_i^{2 / 3}\right\}\right)+\hat{\sigma}_{\mcd}\left(\left\{\bfrd_i^{2 / 3}\right\}\right) \Phi^{-1}(0.99)\right)^{3 / 2} 
\end{equation}
and define a new subset $H^*$ as those $i$ with $\bfrd_i \leqslant c_{\rd}$ but not in $I_{r, \DDC}$. The complementary subset $I \setminus H^*$ defines the new $I_{r,\text{\tiny par}}$. PARAFAC-ALS is then applied to the $\bfX_{H^*}^{(2)}$, while updating its NAs and cellwise outliers in each iteration. This results in the final loading matrices $\bhB$ and $\bhC$. Scores and fitted values for all the observations are computed as in Step 4 (replacing $H_0$ with $H^*$), leading to new updates for their missing values and cellwise outliers that are stored in the imputed data matrix $\btX$.
\item
The final scores and fitted values are obtained as
\begin{align}
\bhA & = \btX (\bhC \odot \bhB)^{\dagger} \label{eq:scores}\\
\bhX & = \bhA (\bhC \odot \bhB)^{\prime} \,. \notag
\end{align}
We thus also impute the outlying cells from rowwise outliers to compute their scores and fitted values. 
We then define the \textit{residual array}
as $\ubR = \ubX - \ubhX$  which has NAs wherever $\ubX$ does. This matrix will be used for outlier detection, outlined in Section~\ref{sect:outlierdetection}. 
Finally, the imputed values for the missing cells in $\btX$ are set to the corresponding cells of $\bhX$.   
\end{enumerate}

Note that the first two steps of MacroPARAFAC are applied to the unfolded matrix $\boldsymbol{X}^{I \times J K}$. This enables us to utilize the DDC algorithm which is a very powerful method to detect cellwise outliers in high-dimensional data, whereas the multivariate outlyingness \eqref{eq:outl} is very appropriate to find a subset of observations that does not include rowwise outliers. We can not compute the outlyingness on the original matrix $\boldsymbol{X}^{I \times J K}$ since all rows could be contaminated by cellwise outliers, hence we compute it on the cell-imputed matrix $\bfX^{(0)}$. From Step 3 on we fully deploy the multiway structure of the data.

\section{Simulation study}
\label{sect:simul}
To evaluate the efficiency and robustness of our proposed method, we compared its performance with several other approaches under different contamination scenarios. The simulation study includes rowwise outliers, cellwise outliers, missing values, and combinations of the three.

The simulation setting for generating clean data $\underline{\boldsymbol{X}}$ is similar to the simulation study in \cite{Engelen:CombPar}. Three-way data are generated that attempt to mimic fluorescence data. 
Specifically, the number of components is set to $F = 2$.  The $\boldsymbol{B}$ and $\boldsymbol{C}$-loadings, of dimension $(76 \times 2)$ and $(61 \times 2)$ respectively, are generated using a mixture of Gaussian distributions:
\begin{equation*}
\frac{1}{3} N(\mu_1, \sigma_1^2)+\frac{1}{3} N(\mu_2, \sigma_2^2)+\frac{1}{3} N(\mu_3, \sigma_3^2)
\end{equation*}
with for each $\ell = 1,2,3$ the center
$\mu_{\ell}$ and variance $\sigma_{\ell}^2$ as specified in Table~\ref{table:mixture}. The scores matrix $\boldsymbol{A}$ of dimension $(50 \times 2)$ is generated from a multivariate normal distribution with center $\boldsymbol{\mu} = (10,10)'$ and $\boldsymbol{\Sigma} = \text{diag}(1,2)$. Hence $I=50, J=76$ and $K=61$.

\begin{table}[ht] 
\caption{Parameters specifying the $\boldsymbol{B}$ and $\boldsymbol{C}$-loadings} 
\label{table:mixture}
\centering
\begin{tabular}{cccccccc} 
\hline
  & & $\mu_1$ & $\mu_2$ & $\mu_3$ & $\sigma_1^2$ & $\sigma_2^2$ & $\sigma_3^2$  \\ [0.5ex] %
\hline 
$\boldsymbol{B}$-loadings & $\boldsymbol{b}_1$ & -8 & 0 & 8 & 10 & 12 & 10 \\
& $\boldsymbol{b}_2$ & 25 & 20 & 15 & 4 & 4 & 4  \\
$\boldsymbol{C}$-loadings & $\boldsymbol{c_1}$ & -8 & 0 & 8 & 10 & 10 & 10  \\
& $\boldsymbol{c}_2$ & -15 & -20 & -25 & 6 & 6 & 6  \\ [1ex] 
\hline 
\end{tabular}
\end{table}

\begin{figure}[!ht]
    \centering
		\includegraphics{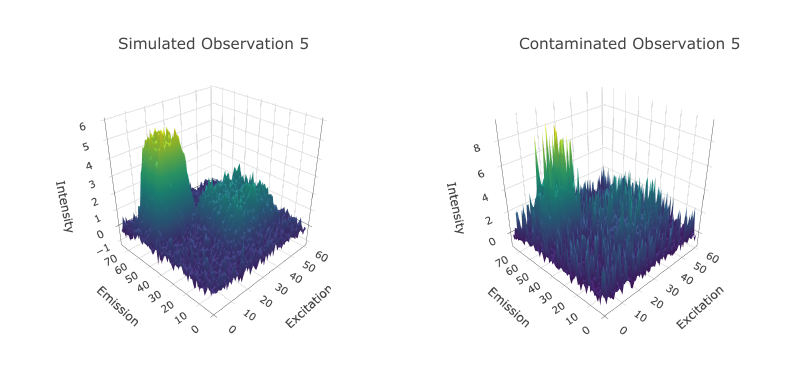}
	     \caption{Landscape from the simulated data without contamination (left) and with $10\%$ cellwise contamination (right).}%
    \label{fig:simuldata_pure_cell}%
\end{figure}

The pure data array $\underline{\boldsymbol{X}}_\text{pure}$ is then defined via
$\boldsymbol{X}_{\text{pure}}^{I \times  JK}=100\, \boldsymbol{A}(\boldsymbol{C} \odot \boldsymbol{B})^{\prime}\, .$
Next noise $\boldsymbol{E}$ is added to the pure data, yielding $\underline{\boldsymbol{X}} = \underline{\boldsymbol{X}}_{\text{pure}} + \underline{\boldsymbol{E}}$. Following \cite{Tomasi:PARAFACnoise} the unfolded noise matrix is defined as:
\begin{equation*}
\boldsymbol{E}^{I \times J K}=\frac{\text {Noise }}{1-\text {Noise}} \left\|\boldsymbol{X}_{\text {pure }}^{I \times J K}\right\|_F \btE^{I \times J K},
\end{equation*}
with $\btE^{I \times J K}$ generated from a normal distribution $N(\boldsymbol{0}, \boldsymbol{\Sigma})$ with $\bSigma=\text{diag}(10)$ and normalised to Frobenius norm of 1. The noise parameter reflects the percentage of noise in the total variance of $\underline{\boldsymbol{X}}$ and is set to 20\%. The left picture of Figure~\ref{fig:simuldata_pure_cell} depicts one landscape from the simulated data array. 

Next the clean data were contaminated with a combination of outliers and missing values. Each setting is characterized through a quartet of percentages $(\rho, \varepsilon_r, \varepsilon_c, \nu)$ that define the generated data. 
First a random set of $\rho I$ horizontal slices of $\ubX$ is selected to remain uncontaminated. Next $\varepsilon_r I$ horizontal slices are randomly drawn from the remaining $(1-\rho)I$ slices and   
all their values $x_{ijk}$ are replaced by $3\, x_{ijk}+1$. This generates rowwise outliers. 
To generate cellwise outliers, we randomly select $\varepsilon_c \,IJK$ cells in the remaining $(1 - (\rho + \varepsilon_r))I$ slices and replace their value $x_{ijk}$ by
$\bar{x}_{.jk} + \gamma \operatorname{std}\left(\boldsymbol{x}_{.jk}\right)$.  Here $\bar{x}_{.jk}$ and $\operatorname{std}\left(\boldsymbol{x}_{.jk}\right)$ are the empirical mean and standard deviation of the column $\boldsymbol{x}_{.jk} =  (x_{1jk},x_{2jk},\ldots,x_{Ijk})'$. Placing all outlying values in the same direction creates a very adversial setting, typically considered to study the effect of cellwise outliers, see e.g.\ \cite{Rousseeuw:DDC}, \cite{Hubert:MacroPCA}, \cite{Raymaekers:cellMCD}. Finally missing values are introduced by putting NA in a random set of $\nu \, I J K$ cells, in every possible row but not in one of the outlying cells.
The right picture of Figure~\ref{fig:simuldata_pure_cell} shows  a simulated landscape with $10\%$ of cellwise outliers and $\gamma = 7$. 

To comprehensively assess the performance of MacroPARAFAC, we
considered many different settings with varying values of $[\rho, \varepsilon_r, \varepsilon_c, \nu]$.  
We report the results for a representative set of contamination scenarios, with $\gamma = 7$. First we consider four settings without missing values:
\begin{itemize}
    \item $[1,0,0,0]$: uncontaminated data (U) 
    \item $[0.8,0.2,0,0]$: 20\% of rowwise outliers (R20)
    \item $[0,0,0.1,0]$: 10\% of cellwise outliers (C10)
    \item $[0,0.1,0.1,0]$: 10\% of rowwise outliers and 10\% of cellwise outliers (R10C10)
\end{itemize}
Note that in the last two settings cellwise outliers are generated without leaving any of the rows uncontaminated ($\rho=0)$. This creates a very adversarial setting in which all the observations are affected. We can then expect rowwise robust methods to break down. 
We also consider data with missing values:
\begin{itemize}
    \item $[0,0,0, 0.2]$: 20\% of NAs (NA20)
    \item $[0,0.1,0.1,0.1]$:  10\% of rowwise outliers, 10\% of cellwise outliers and 10\% of NAs (R10C10NA10)
    \item $[0,0.1,0.1,0.2]$: 10\% of rowwise outliers,  10\% of cellwise outliers and 20\% of NAs (R10C10NA20)
    \item $[0.4,0.1,0.1,0.1]$: 40\% of uncontaminated rows, 10\% of rowwise outliers, 10\% of cellwise outliers and 10\% of NAs (U40R10C10NA10). 
\end{itemize}
The last setting implies that the cellwise contamination is spread over 50\% of the rows, hence the proportion of contaminated cells in those rows is on average 20\%, creating overall a very challenging situation. 

For the settings without missing values, we evaluate four methods:
\begin{enumerate}
    \item PAR: classical PARAFAC using the function `parafac' from the R package \texttt{multiway}. It uses random starts for the $\bB$ and $\bC$ loadings. We did not specify any constraint for the loadings. 
    \item RPAR: RPARAFAC proposed in \cite{Engelen:robParafac}, available in the R package \texttt{rrcov3way} as the function `Parafac' with option \texttt{robust = TRUE} \citep{R:rrcov3way}. 
    \item DDC-RPAR: a simple alternative to MacroPARAFAC which first detects and imputes cellwise outliers with DDC (as in step 1 of MacroPARAFAC) and then applies RPARAFAC to the imputed data array. 
    \item MacroPAR: MacroPARAFAC.
\end{enumerate}

The simulation study with NAs includes the following methods:
\begin{enumerate}
    \item PAR: classical PARAFAC with iterative updates of the NAs using the (undocumented internal) function `parafac\_3wayna' from the R package \texttt{multiway}. This function uses 10 different random initialisations for the missing values and selects the fit with smallest loss function. 
    \item DDC-RPAR: the missing values and the cellwise outliers are imputed with DDC, followed by RPARAFAC. 
    \item  MacroPAR: MacroPARAFAC.
\end{enumerate}
The robust methods RPAR, DDC-RPAR and MacroPAR are run with $h = 39 = \lceil 0.75(I+1) \rceil$.

To evaluate the different methods, we compute the \textit{mean squared error} of the regular cells:
\begin{equation} \label{eq:MSE}
    \text{MSE}=\frac{1}{m} \sum_{i=1}^I \sum_{j=1}^J \sum_{k=1}^K w_{ijk}\left(x_{i j k}-\hat{x}_{i j k}\right)^2
\end{equation}
with $w_{ijk} = 0$ if $i$ corresponds to the set of generated rowwise outliers or if the cell $x_{ijk}$ is generated as a cellwise outlier or as a missing value. For all other cells $w_{ijk} = 1$, and $m = \sum_{ijk} w_{ijk}$. 
Consequently, the MSE exclusively involves regular cells, providing a robust measure of model performance.
Both for the $\bB$ and $\bC$ loadings, we also measure the angle between the estimated subspace spanned by the loadings and the generated subspace. This angle is defined as the second principal angle between the two subspaces \citep{Hubert:ROSPCA} and computed with the \texttt{subspace} function from the R package \texttt{pracma} \citep{R:pracma}.
Obviously, we want all our diagnostics to be as small as possible. Boxplots of the results are shown based on 100 replicates.

\begin{figure}[!htb]   
\centerline{%
\includegraphics[height=.25\paperheight]{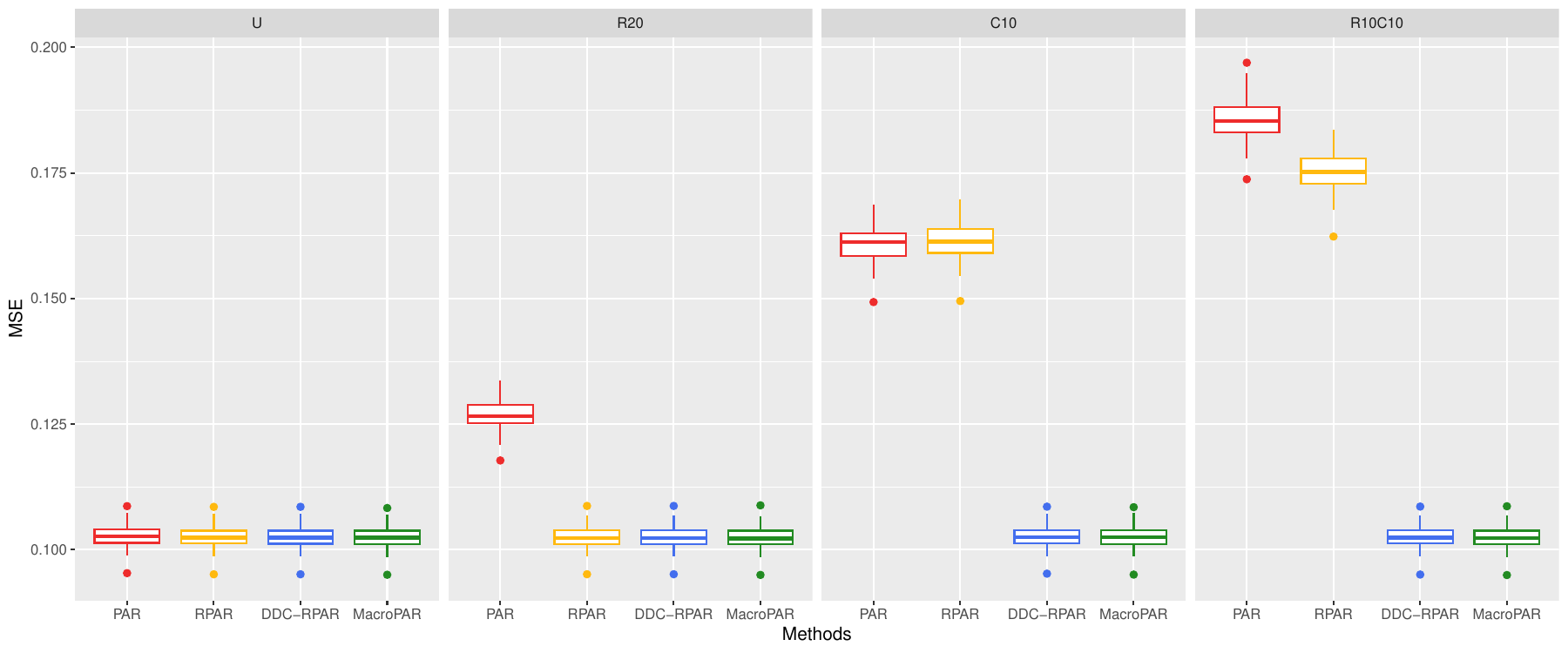}%
}%
\caption{Boxplots of MSE for simulated data under varying contamination schemes without missing values.}
\label{fig:simul_boxplotMSE_noMISS}
\end{figure}

\begin{figure}[!htb]   
\centerline{%
\includegraphics[height=.25\paperheight]{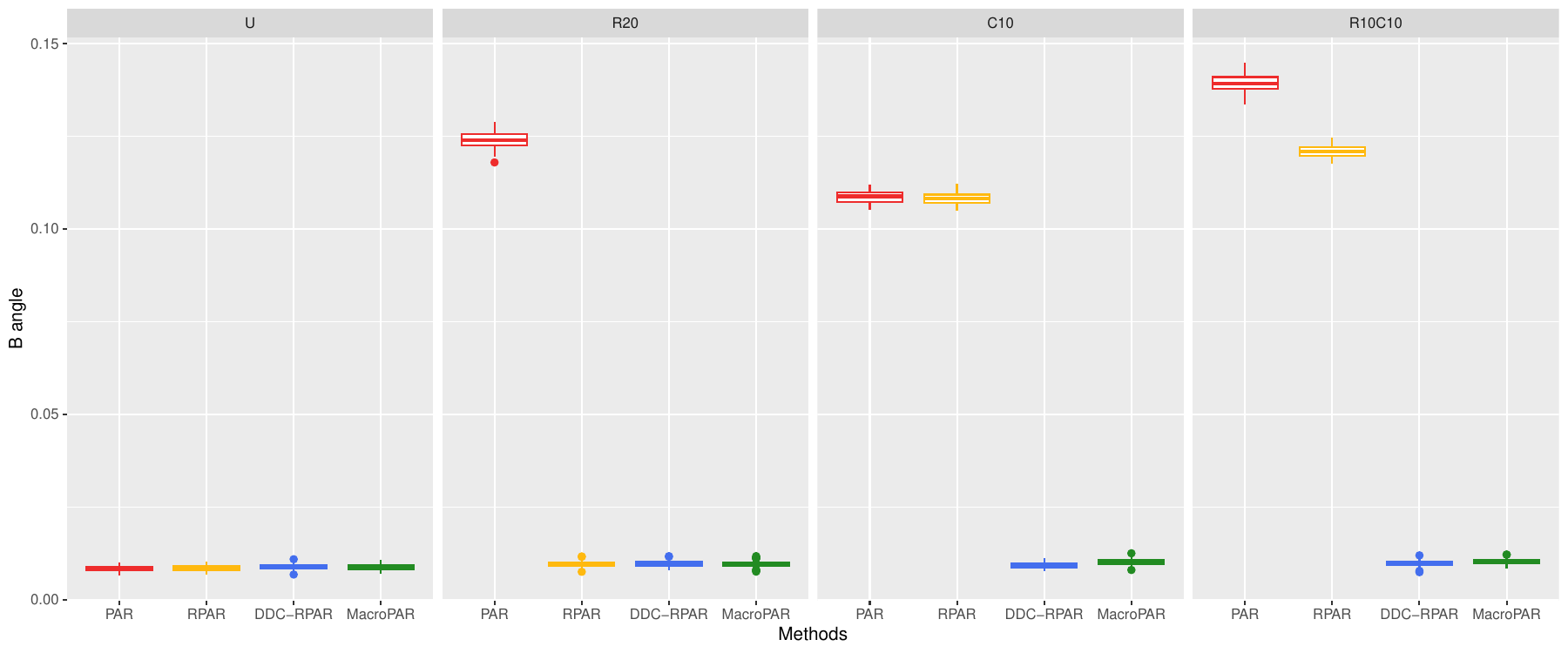}%
}%
\caption{Boxplots of $\bB$-angle for simulated data under varying contamination schemes without missing values.}
\label{fig:simul_boxplotBangles_noMISS}
\end{figure}

In Figure~\ref{fig:simul_boxplotMSE_noMISS} and Figure~\ref{fig:simul_boxplotBangles_noMISS} 
the results for the MSE and the angle of the $\bB$-loadings ($\bB$-angle) are presented for the simulations without NAs. The angles of the $\bC$-loadings were very similar and are therefore not included. 
The supplementary material contains the same boxplots truncated at MSE values above 0.1075 and $\bB$-angles above 0.015 to better visualize the differences between the well-performing methods (Supplementary Figures~\ref{fig:simul_boxplotMSE_noMISS_trunc} and~\ref{fig:simul_boxplotBangles_noMISS_trunc}).

In the absence of contamination (U) all four methods behave similarly. The impact of 20\% of rowwise contamination (R20) is clearly visible in the second panel. RPARAFAC, DDC-RPAR and MacroPARAFAC can handle this situation well, while PARAFAC experiences a noticeable increase in MSE and a considerable increase in the $\bB$-angle. 
Under cellwise contamination (C10) and both rowwise and cellwise contamination (R10C10), PARAFAC and RPARAFAC break down completely (as expected), whereas the DDC-RPAR and MacroPARAFAC estimates are not highly affected. 

Simulation results for data with NAs are reported in 
Figure~\ref{fig:simul_boxplotMSE_MISS}
and Figure~\ref{fig:simul_boxplotBangles_MISS} (and in Supplementary Figure~\ref{fig:simul_boxplotMSE_MISS_trunc}
and Figure~\ref{fig:simul_boxplotBangles_MISS_trunc}).
PARAFAC, DDC-RPAR and MacroPARAFAC show comparable results in the presence of missing values only (NA20). Once rowwise and cellwise contamination is added, a discernible increase in the MSE and loading estimates becomes apparent for PARAFAC, whereas the DDC-RPAR and MacroPARAFAC results are quite stable. 

\begin{figure}[!htb]   
\centerline{%
\includegraphics[height=.25\paperheight]{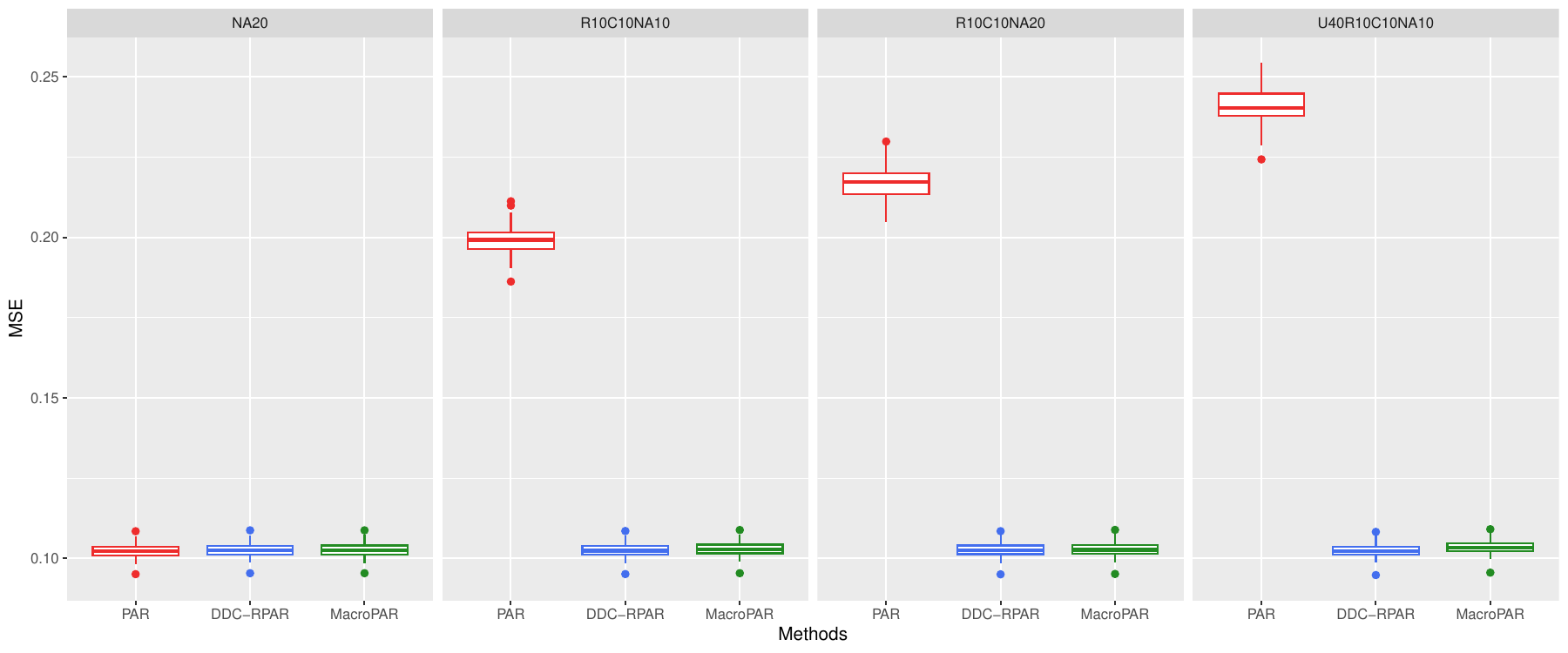}%
}%
\caption{Boxplots of MSE for simulated data under varying contamination schemes and missing values.}
\label{fig:simul_boxplotMSE_MISS}
\end{figure}

\begin{figure}[!htb]   
\centerline{%
\includegraphics[height=.25\paperheight]{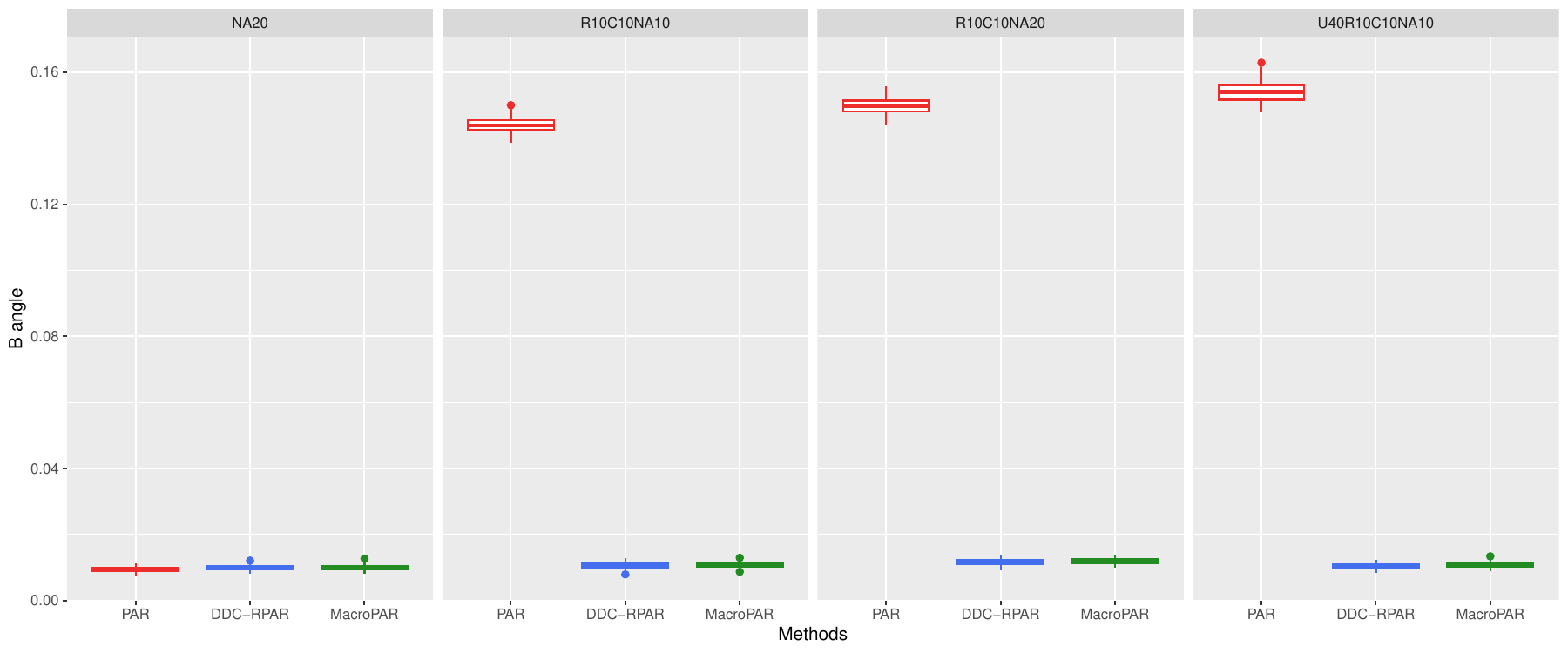}%
}%
\caption{Boxplots of $\bB$-angle for simulated data under varying contamination schemes and missing values.}
\label{fig:simul_boxplotBangles_MISS}
\end{figure}

In the supplementary material we also present average MSE curves (over 10 runs) when $\gamma$ is varied from 0 to 10, for the settings R10C10 and R10C10NA20.
Supplementary Figure~\ref{fig:Simulation_MSEevol} shows that for all methods the MSE is increasing up to $\gamma=3$. This is not surprising as low values of $\gamma$ generate asymmetric inliers. Only for cellwise robust methods the MSE goes down from $\gamma=4$ on. This shape is similar to MSE curves of MacroPCA presented in \cite{Hubert:MacroPCA}.

\begin{figure}[!htb]   
\centerline{%
\includegraphics[scale=0.5]{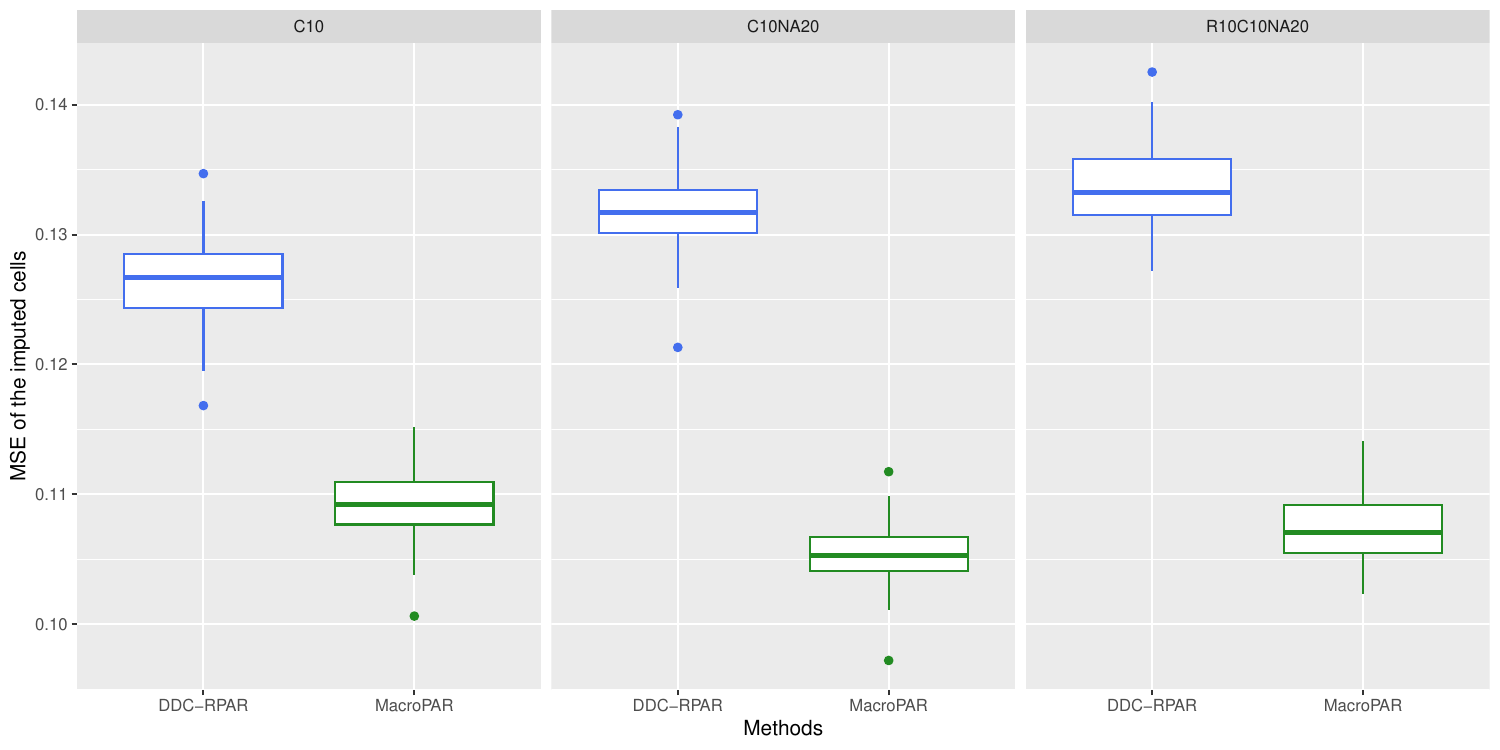}%
}%
\caption{Boxplots of MSE of imputations for simulated data under varying contamination schemes and missing values.}
\label{fig:simul_boxplotMSE_imputed}
\end{figure}

From all the simulation results we notice that the differences between DDC-RPAR and MacroPARAFAC are small. This is not unexpected since DDC is very powerful in detecting cellwise outliers, whereas RPARAFAC is highly resistant towards rowwise outliers. 
To study both methods in more detail, we finally also compare their imputed values in the missing and contaminated cells. More precisely, we compute 
\begin{equation*}
    \text{MSE}_\text{imp}=\frac{1}{IJK-m} \sum_{i=1}^I \sum_{j=1}^J \sum_{k=1}^K (1-w_{ijk})\left(x_{i j k}-\tilde{x}_{i j k}\right)^2
\end{equation*}
with $w_{ijk}$ and $m$ as in \eqref{eq:MSE}. For DDC-RPAR it holds that $\tilde{x}_{ijk} = \tilde{x}_{ijk}^{(0)}$ (see step 1 of MacroPARAFAC). 
The resulting boxplots (for $\gamma=7$) are shown in Figure~\ref{fig:simul_boxplotMSE_imputed}. We conclude that iteratively updating the NAs and cellwise outliers, as done in MacroPARAFAC, is beneficial. This is not surprising since the imputed values of DDC are computed from the matricized data array ignoring its intrinsic multiway structure. MacroPARAFAC on the other hand estimates the imputed values through the multiway PARAFAC model.

The supplementary material contains a study of the computing time of MacroPARAFAC. We considered  $I \in \{20, 125, 300, 500\}$, $J \in \{ 76, 152\}$, $K \in \{61, 122\}$ and ran MacroPARAFAC 10 times (on a MacBook Pro 15.6 with Apple M3 Pro chip and 18 GB memory). The average computing times of the full algorithm are displayed in Supplementary Figure~\ref{fig:Simulation_MacroPAR_CompTime} with full lines, whereas the dotted lines depict the computing time of DDC which is applied to the unfolded array.
The results show that it is still feasible to compute MacroPARAFAC at larger data sizes. The DDC part accounts for roughly 75\% of the overall computing time. Despite the computationally efficient O$(IJK \log(IJK))$ time complexity of FastDDC, its computation may become problematic at ultrahigh dimensional data. In those settings a computationally less demanding cellwise robust method could be used, such as the wrapping method proposed by \cite{Raymaekers:FastCorr}, which only operates on the marginal distributions and does not consider the correlations among the variables. 

\section{Outlier detection}
\label{sect:outlierdetection}

A robust method, such as MacroPARAFAC, does not only provide robust estimates, it also allows for outlier detection. 
First, we can identify outlying cells in $\ubX$ as those cells whose absolute standardized residual $|r_{ijk}|/\hat{\sigma}_{jk}$ exceeds a high quantile of the $\sqrt{\chi^2_1}$ distribution. We estimate the scales $\sigma_{jk}$ by a robust M-scale applied to the columns $\br_{.jk}$, yielding $\hat{\sigma}_{jk}$ for all $j=1,\dots,J$ and $k=1,\dots,K$. In our examples we use as cutoff value $c_r = \sqrt{\chi^2_{1,0.998}} = 3.09$.  

To visualize the outlying cells we can then make residual maps as in \cite{Hubert:MacroPCA} for two-way data. Such a residual map is essentially a heatmap which colors each cell according to its standardized residual. Regular cells are colored yellow, cells with a positive standardized residual $ > c_r$ receive a color which ranges from light orange to red, whereas negative standardized residuals $ < - c_r$ are colored from light purple to dark blue. Missing values are set to white. 

\begin{figure}[!ht]   
\centerline{%
\includegraphics{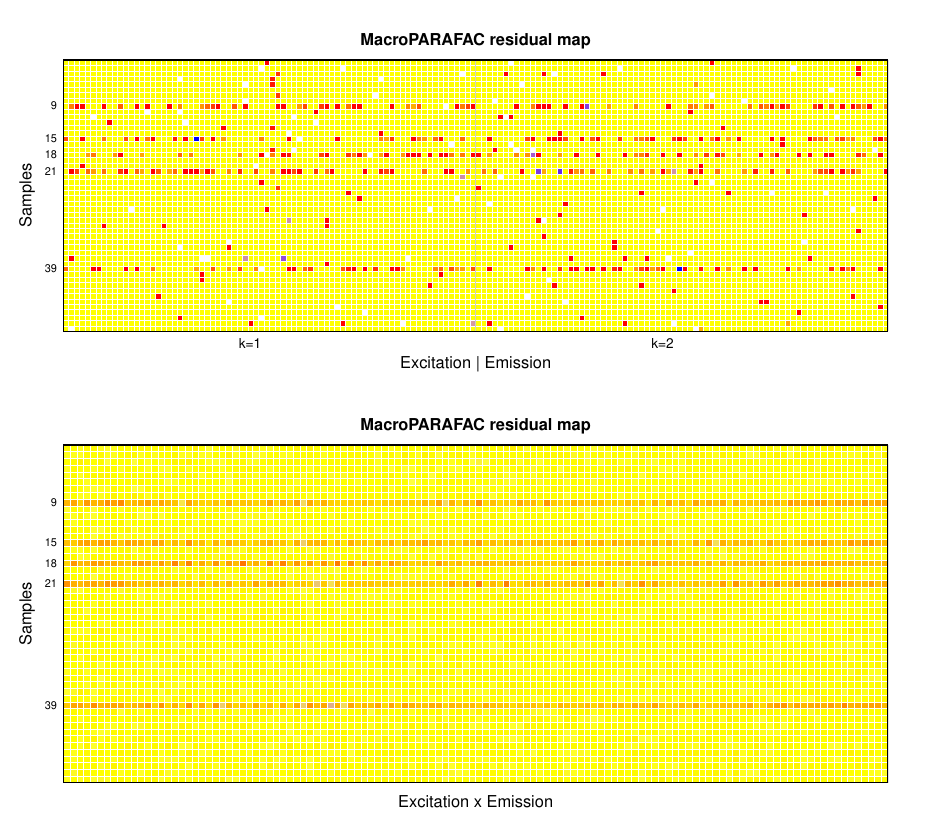} 
}
\caption{Residual maps of a generated data set with a selection of columns (upper) and with aggregated columns (lower).}
\label{fig:simul_ResidMap}
\end{figure}

In our multiway setting, we can matricize the residual array (typically along the first mode) to make a residual map that shows the cells of all the observations. As this might become a very wide map when $J$ and/or $K$ is large, it is possible to select a subset of the columns, or to aggregate the cells into blocks. Figure~\ref{fig:simul_ResidMap} shows residual maps of one of our generated data sets with 10\% of rowwise outliers, 1\% of cellwise outliers, and 1\% of missing values. In the upper panel we have selected the first 152 columns of the unfolded array, which are the cells with $k=1$ followed by those with $k=2$. In the lower panel the color of each cell is the average color of 38 consecutive columns of the unfolded residual array. The first two columns thus correspond with $k=1$, the second two with $k=2$ and so on. The rowwise outliers are easily spotted as they have many outlying cells. Note that almost all outlying residuals are positive as the data were generated as such. 

Whereas the residual maps show information at the cell level, we can also quantify the degree of outlyingness at the sample level. For this we consider an enhanced version of the outlier map proposed in \cite{Hubert:ROBPCA} and \cite{Engelen:robParafac}. This outlier map shows for each observation how well it fits the model. On the vertical axis it displays the residual distance, defined as 
\begin{equation}
    \RD_i = \| \bcX_i - \bhX_i \|_F 
\end{equation}
whereas the horizontal axis shows the score distance 
\begin{equation}
    \SD_i = \sqrt{ (\ba_i - \hat{\boldsymbol{\mu}}_{\ba})' \bhSigma_{\ba}^{-1} (\ba_i -\hat{\boldsymbol{\mu}}_{\ba}) }
\end{equation}
with $\hat{\boldsymbol{\mu}}_{\ba}$ and $\bhSigma_{\ba}$ the MCD location and scatter estimates of the scores \citep{Hubert:WIRE-MCD2}.
Cutoff values for both distances are given by $c_{\rd}$ defined in \eqref{eq:cutoff_rd} and $c_{\sd} = \sqrt{\chi^2_{F,0.998}}$ (under the assumption that the regular scores are Gaussian) and visualized with dashed red lines. 

\begin{figure}[!ht]
\begin{center}
    \includegraphics[scale=0.5]{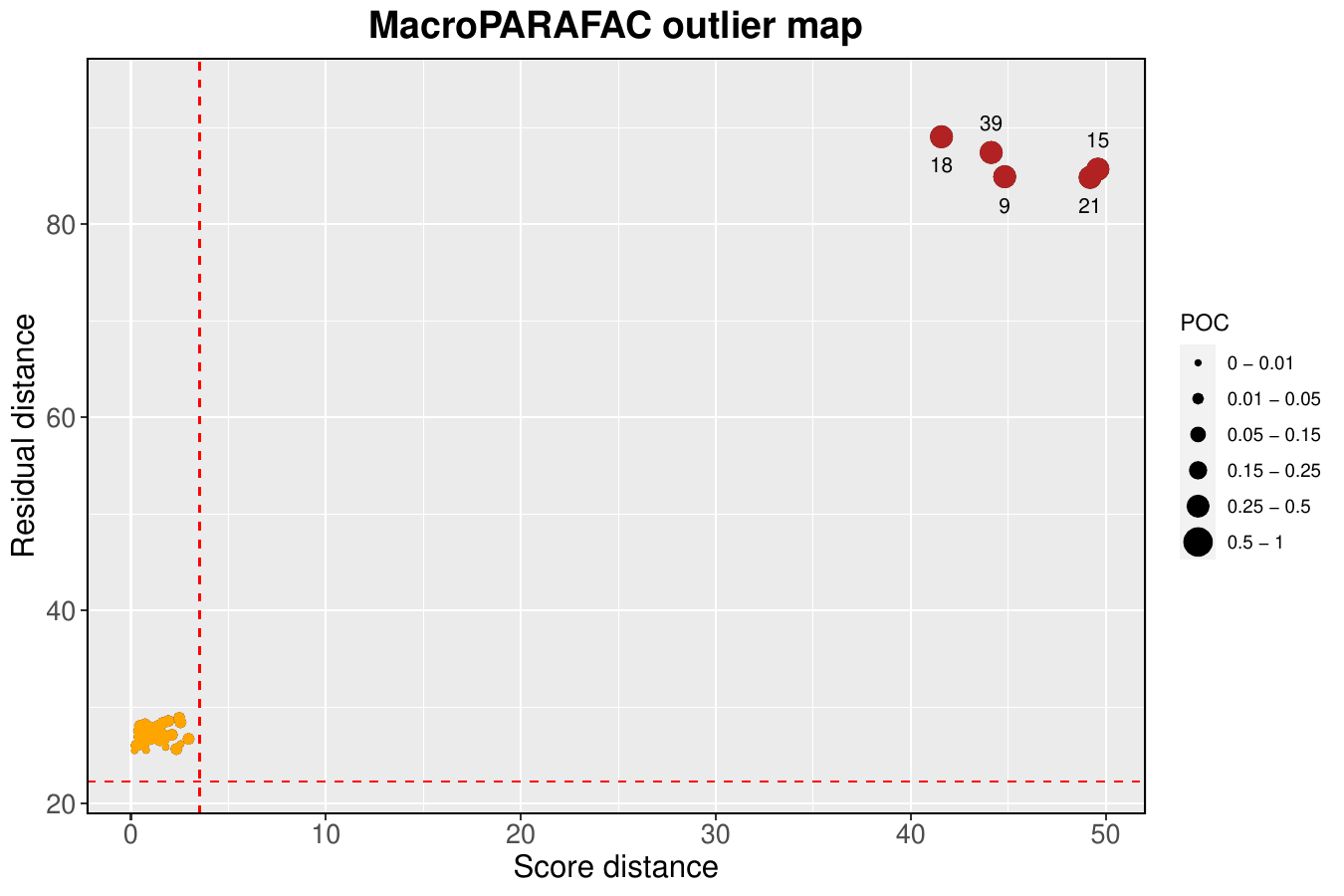}
\end{center}
\vspace{-3mm}
\caption{Enhanced outlier map of the data from Figure~\ref{fig:simul_ResidMap}.}
\label{fig:simul_OutlierMap}
\end{figure}

One could argue that observations with outlying $\RD_i$ or $\SD_i$ can be classified as rowwise outliers (generated from another distribution than the majority), but this is not necessarily true as also observations with one or a few outlying cells can expose a very large RD or SD.  
Figure~\ref{fig:simul_OutlierMap} shows the enhanced outlier map for the generated data of Figure~\ref{fig:simul_ResidMap}. We notice that all observations have an outlying residual distance, also the group with a regular score distance. This is because all cases have outlying cells. 

Making a formal distinction between rowwise outliers on one hand and observations with some outlying cells on the other hand is often not feasible, see also the discussion in \cite{Raymaekers:Challenges}. We can however consider several diagnostics to guide us in classifying the observations. First, we can compute for each observation its Percentage of Outlying Cells (POC), defined as
\begin{equation*}
    \text{POC}_i = \frac{1}{JK}\sum_{j=1}^J \sum_{k=1}^K I(|r_{ijk}|/\hat{\sigma}_{jk} > c_r) \,.
\end{equation*}
Obviously, the larger $\text{POC}_i$ is, the more probable  that the observation deviates from the regular pattern of the observations. On the enhanced outlier map the size of the points is plotted according to their POC. In Figure~\ref{fig:simul_OutlierMap} we clearly notice the difference between the observations with few outlying cells (with a POC that is on average equal to 1.2\%) and the generated rowwise outliers (whose POC is on average 39\%). 

As a secondary diagnostic, we assess the impact on the fit of each observation by imputing its outlying cells.
We first compute for each case its  \textit{imputed residual distance}
\begin{equation}
   \tRD_i = \| \btX_i - \bhX_i \|_F
\end{equation}
and then color the points on the outlier map according to the following rules:
\begin{itemize}
    \item green points have both $\RD_i < c_\rd$ and $\tRD_i < c_\rd$
    \item orange points have $\RD_i > c_\rd$ and $\tRD_i < c_\rd$
    \item red points have $\RD_i > c_\rd$ and $\tRD_i > c_\rd$
\end{itemize}
Green-colored observations can be interpreted as regular ones, orange-colored observations become regular when their cellwise outliers are imputed, whereas red-colored observations do not yet fit well even when their outlying cells are imputed. For our generated data set with 5 rowwise outliers and 45 observations with outlying cells, it turns out that the imputed residual distance of the rowwise outliers is still larger than what we expect for regular observations. Hence the color on the enhanced outlier map nicely discriminates between rowwise outliers and observations with some outlying cells. 

\begin{figure}[!ht]
\begin{center}
    \includegraphics[scale=0.55]{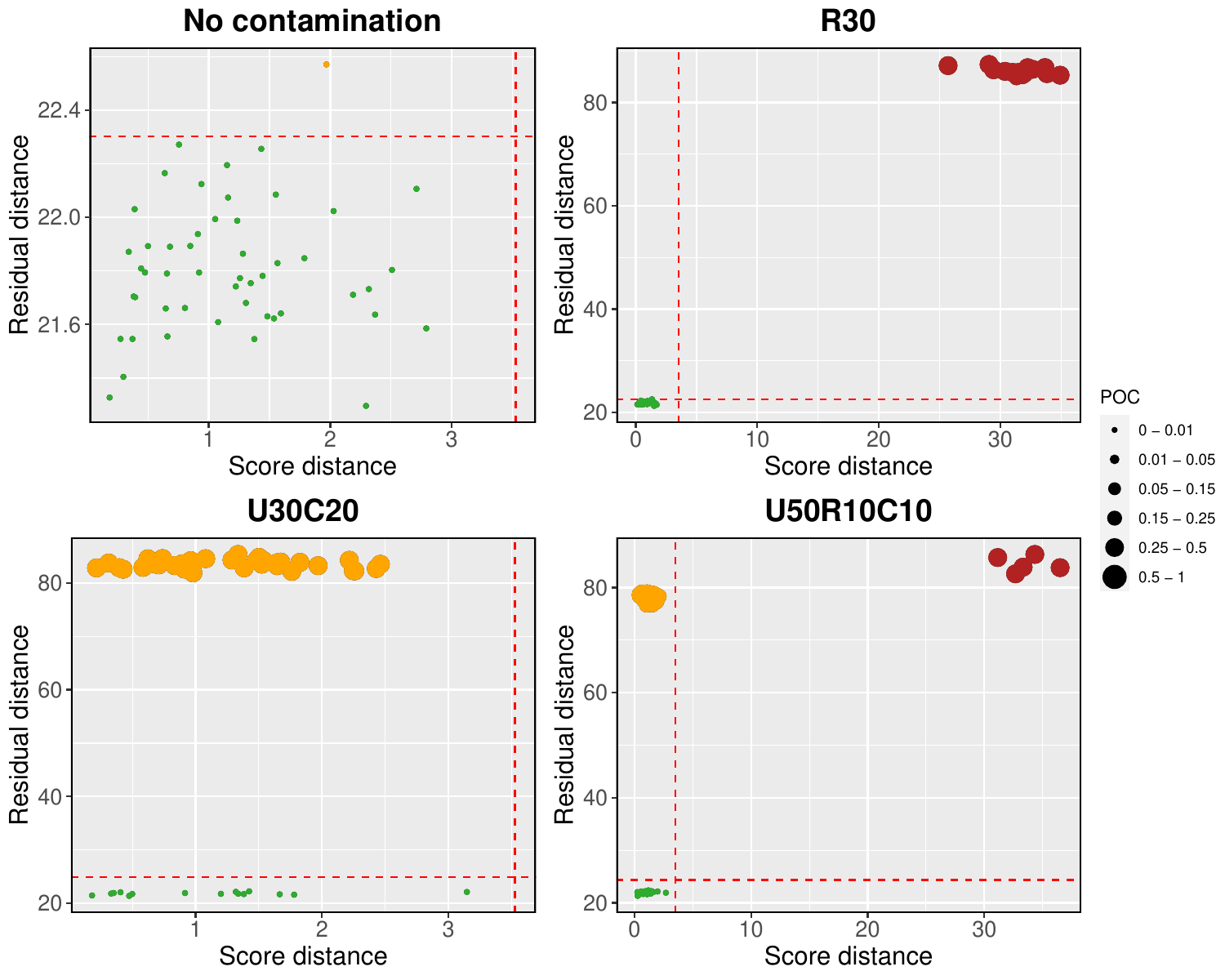}
\end{center}
\vspace{-3mm}
\caption{Enhanced outlier map of four generated data sets.}
\label{fig:simul_OutlierMap_4}
\end{figure}

In Figure~\ref{fig:simul_OutlierMap_4} we illustrate this enhanced outlier map on several of our generated data sets, defined through the quartet $[1,0,0,0]$ (no contamination), $[0.7,0.3,0,0]$ (30\% of rowwise outliers, R30), $[0.3,0,0.2,0]$ (30\% of uncontaminated rows, 20\% of cellwise outliers, U30C20) and $[0.5,0.1,0.1,0]$ (50\% of uncontaminated rows, 10\% of rowwise outliers, 10\% of cellwise outliers, U50R10C10). We see a nice correspondence between the size and the color of the observations and their generating process. 

\section{Real data example}
\label{sect:realdata}

To further evaluate the effectiveness of our proposed algorithm and outlier detection diagnostics, we analyse the Dorrit data, a laboratory-generated fluorescence data set consisting of $I = 27$ different mixtures of $F=4$ fluorophores (hydroquinone, tryptophan, phenylalanine and dopa). For each mixture, its Excitation Emission Matrix (EEM) is measured with emission wavelengths from 250 nm to 482 nm at 2 nm intervals, and excitation wavelengths from 230 nm to 315 nm at 5 nm intervals. This yields a  $(27 \times 116 \times 18)$ data array. 
Using the pure samples measured for each compound, the excitation and emission profiles for each compound are obtained \citep{Baunsgaard:dorrit}. These pure loadings serve as a reference.  Samples 2, 3 and 5 are very anomalous observations. They have very high concentrations for some fluorophores, resulting in censored values (NAs) at the largest intensities. Previous casewise robust analysis has pointed out that samples 4, 10, 12 and 16 also seem to face some unusual characteristics \citep{Riu:RobParJack, Hubert:RParafac-SI}. 

\begin{figure}[!ht]
    \centering
    \includegraphics{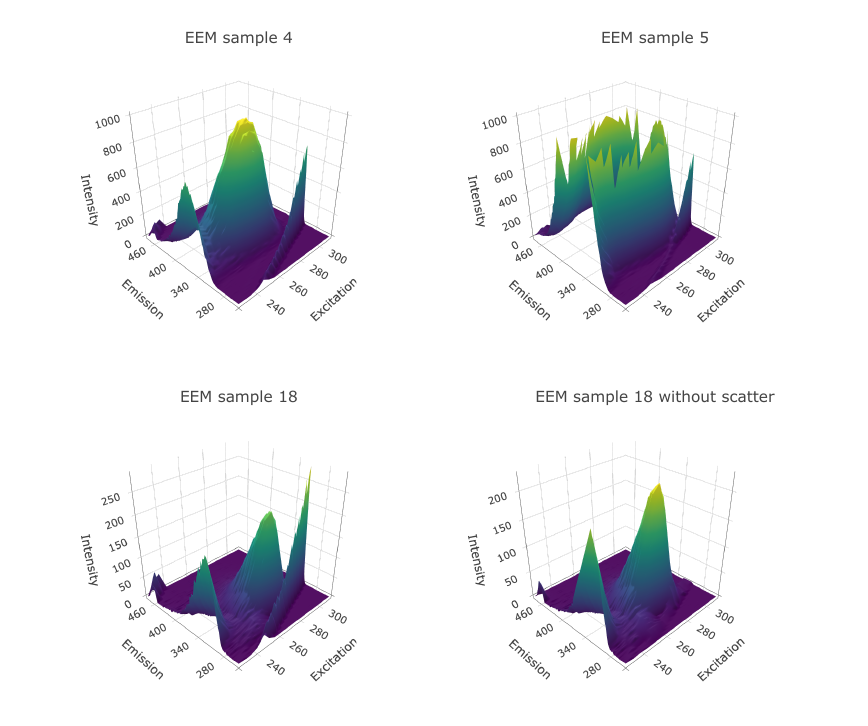}
    \caption{Different landscapes of the original Dorrit dataset, and after removal of the Rayleigh and Raman scattering.}%
    \label{fig:Dorrit_landscapes}%
\end{figure}

The landscapes of samples 4, 5 and 18 are shown in  Figure~\ref{fig:Dorrit_landscapes}. 
The rounded tops represent the fluorophores, whose heights are proportional to the concentrations present in that sample. We also notice some sharp peaks along certain diagonals, caused by Rayleigh and Raman scattering. This affects all samples at specific excitation-emission wavelengths. 
Such a situation cannot be handled well by our method, since a variable with large values for all samples shows a consistent pattern of large measurements, hence they are not flagged as outlying values. Aberrant behaviour within a variable can only be detected when a majority of samples has regular values for it, or in this case, contains no scattering. 
To mitigate this issue, we follow the automated scattering removal method outlined in \cite{Engelen:Scatter} and replace the scattering in all the samples with interpolated values. 
The result for sample 18 is visualized in the lower right panel of Figure~\ref{fig:Dorrit_landscapes}.

\begin{figure}[!ht]
\centering
    \includegraphics{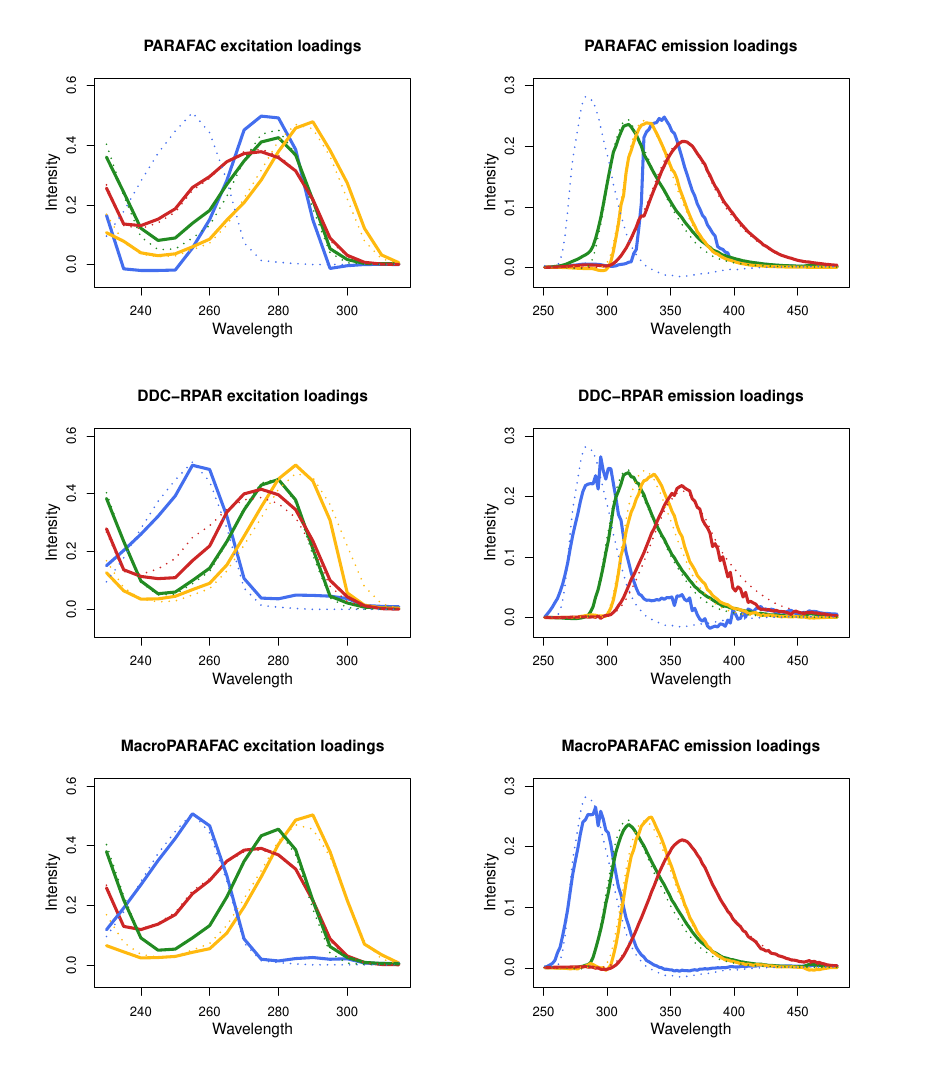}
\caption{Excitation (left) and emission (right) spectra from the Dorrit data. The dotted curves are the pure loadings of phenylanlanine (blue), dopa (green), hydroquinone (yellow) and tryptophan (red). Full curves are the PARAFAC estimates (first row), DDC-RPAR estimates (second row) and MacroPARAFAC estimates (third row).}
\label{fig:Dorrit_loadings_noscatter}
\end{figure}

We first apply PARAFAC (for data with missings), DDC-RPAR and MacroPARAFAC (with $h=21 = \lceil 0.75(I+1) \rceil$) to the preprocessed data. The resulting excitation and emission  loadings (columns of $\bhB$ and $\bhC$) are depicted in Figure~\ref{fig:Dorrit_loadings_noscatter}. In each figure the dotted curves correspond with the pure loadings. 
The PARAFAC loading vectors for phenylanlanine (blue-colored) are clearly highly distorted, whereas the MacroPARAFAC estimates are quite well aligned with the pure loadings. The DDC-RPAR loadings yield intermediate results. 
The distorted PARAFAC estimates are due to the outlying samples 2, 3 and 5. If they are removed from the analysis, the resulting loadings (not shown) look very similar to the pure spectra.    

\begin{figure}[!ht]
\begin{center}
    \includegraphics[scale=0.5]{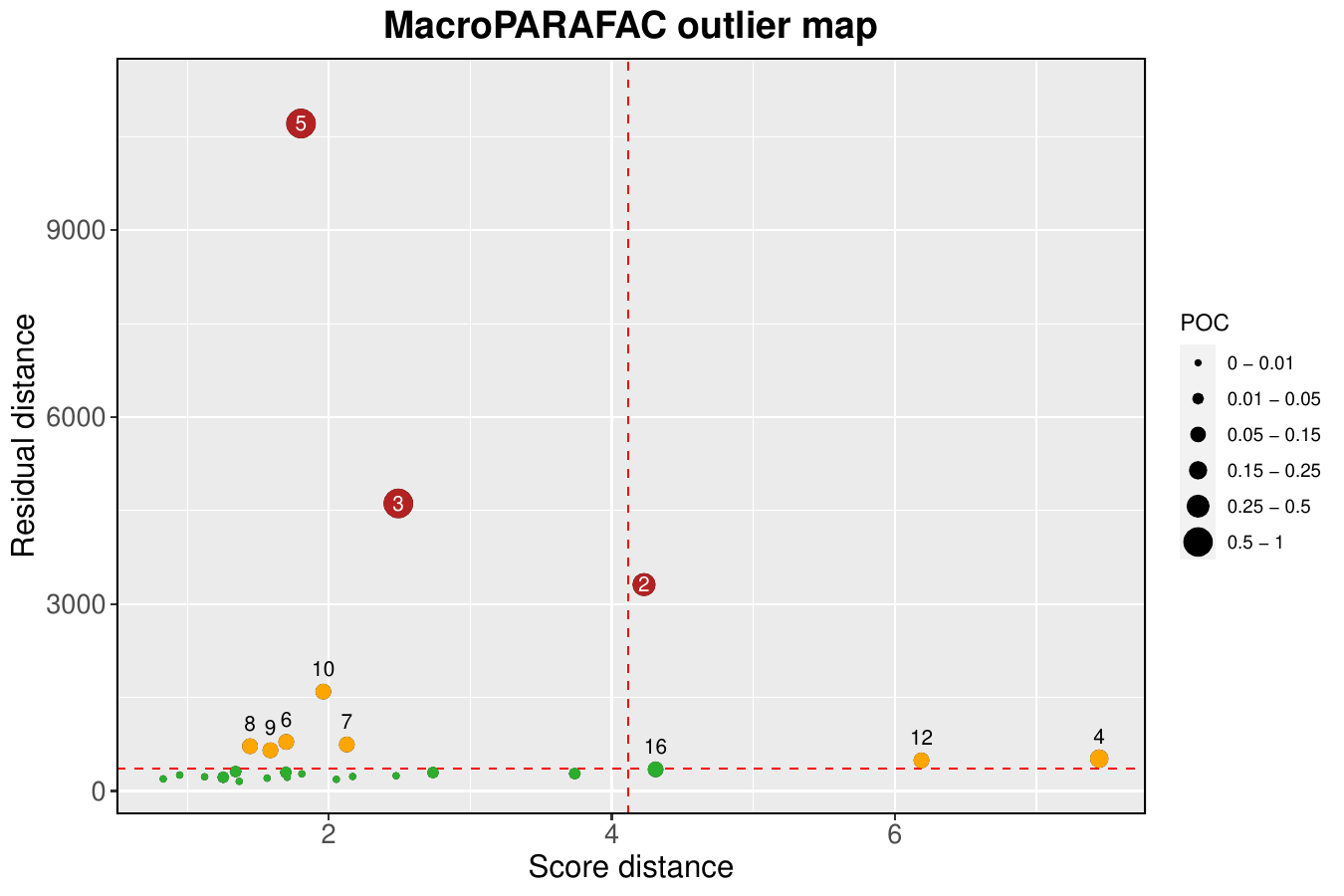}
\end{center}
\vspace{-3mm}
\caption{Enhanced outlier map of the Dorrit data.}
\label{fig:Dorrit_OutlierMap_noscatter}
\end{figure}

Next we look at several outlier diagnostics.
The enhanced outlier map obtained with MacroPARAFAC is shown in Figure~\ref{fig:Dorrit_OutlierMap_noscatter}. The three samples (2, 3 and 5) clearly stand out with an unusual large RD. Their size and color also indicate that they have more than 25\% of outlying cells, and an abnormally large imputed RD. They could be classified as rowwise outliers. Five observations (6, 7, 8, 9 and 10) have around 10\% of outlying cells and a somewhat larger RD that becomes regular when the outlying cells are imputed. Their scores are estimates of the concentrations of the fluorophores in the mixture, and don't look suspicious. Samples 4 and 12 are mostly characterized by having a large SD. Also sample 16 somewhat stands out. This is all in agreement with earlier studies. 
Note however that outlier maps based on RPARAFAC and RPARAFAC-SI yielded very large SD values for samples 2, 3 and 5, classifying them as bad leverage points. This difference is due to a different computation of the scores. Here we estimate the scores following Equation \eqref{eq:scores}, hence based on the data with imputed values for both the NAs and the outlying cells. In \cite{Engelen:robParafac} and \cite{Hubert:RParafac-SI} the scores
were computed using imputed values for the missing values only. 

\begin{figure}[!ht]
\begin{center}
    \includegraphics[scale=0.5]{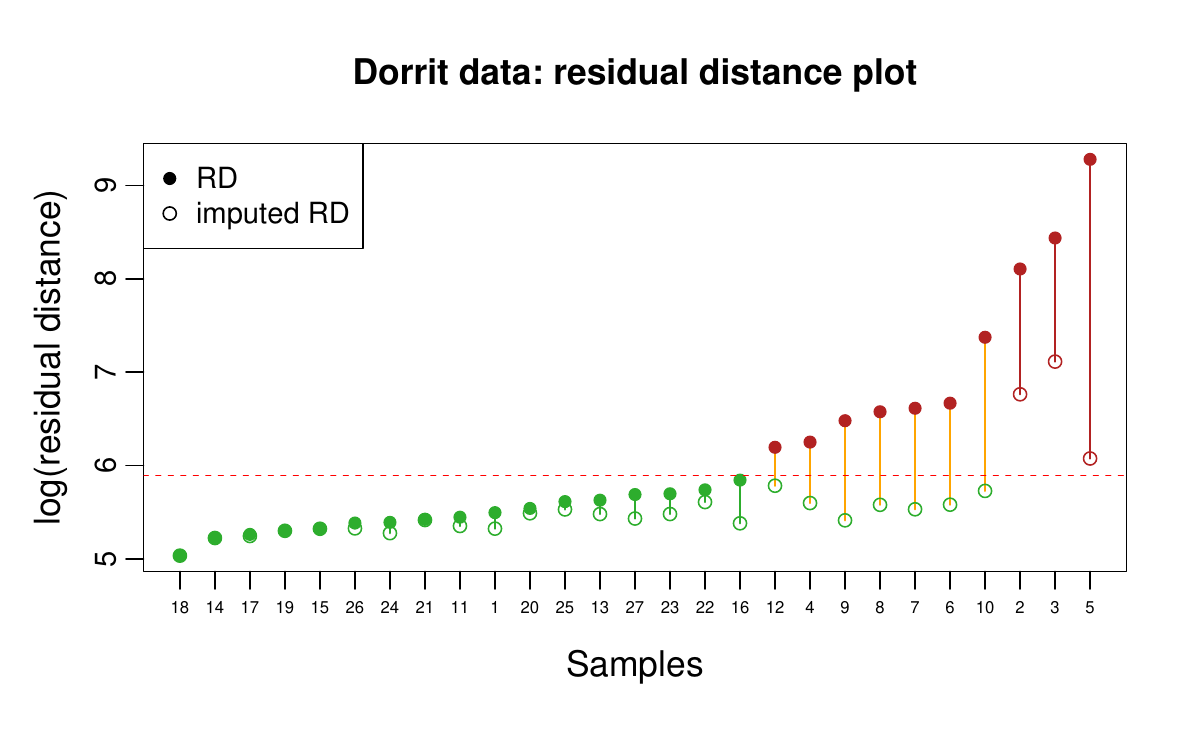}
\end{center}
\vspace{-7mm}
\caption{Plot of the MacroPARAFAC residual distances and imputed residual distances of the Dorrit data.}
\label{fig:Dorrit_ResidualDistanceMap_noscatter}
\end{figure}

The outlier map does not show the imputed residual distance to avoid ending up with an overloaded plot. To provide more insight into the effect of imputation on the residual distance, we show in 
Figure~\ref{fig:Dorrit_ResidualDistanceMap_noscatter}
how for each sample its RD (with filled dot) is reduced to its imputed RD (with open circle). The results are plotted on the log scale and the samples are ranked following increasing RD. Values above $c_\rd$ are red colored, values below this cutoff receive a green color. The segments are colored following the colors defined at the outlier map. Although sample 5 is most outlying in terms of RD, samples 2 and 3 have a substantially larger imputed RD. 

\begin{figure}[!ht]
\begin{center}
    \includegraphics[scale=0.9,trim=0cm 0cm 0cm 1cm, clip]{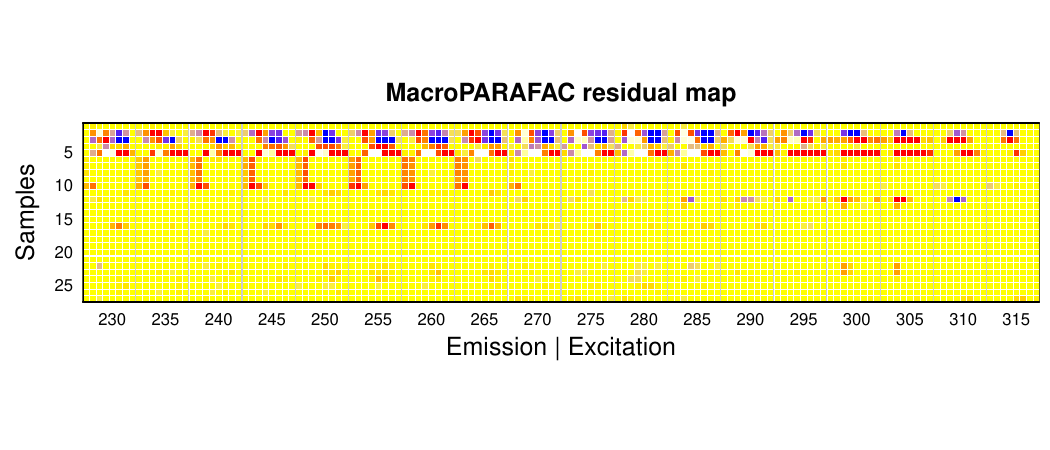}
\end{center}
\vspace{-15mm}
\caption{MacroPARAFAC residual map of all samples matricized according to the first mode.}
\label{fig:Dorrit_ResidualMap_noscatter}
\end{figure}

To further investigate the root cause of the outlying samples  we can look at the residual map of all the observations, matricized according to the first mode, see Figure~\ref{fig:Dorrit_ResidualMap_noscatter}.
Here the cells in the data array are aggregated such that the first 8 columns in the residual map correspond with the emission wavelengths at  excitation wavelength  230 nm ($k=1$), the next 8 columns at excitation wavelength  235 nm ($k=2$) and so on. This yields 18 blocks, separated by thin grey lines.  
We immediately spot the highly outlying samples 2, 3 and 5 as they have many outlying cells. Also other samples with an outlying behavior (4, 6, 7, 8, 9, 10, 12 and 16) can be distinguished. In particular we notice that samples 6 - 10 show a similar pattern. 

\begin{figure}[!ht]
\begin{center}
    \includegraphics[scale=0.5]{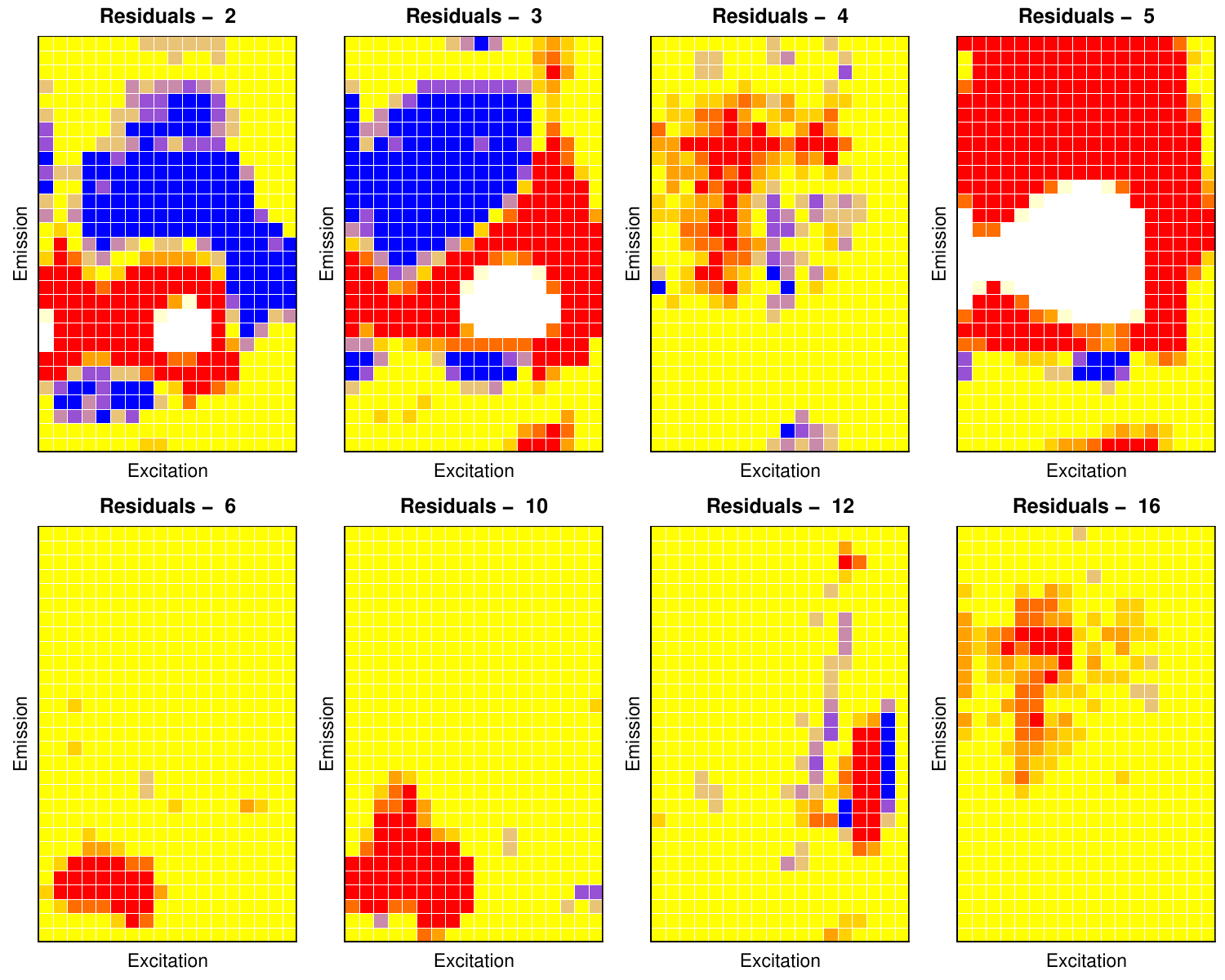}
\end{center}
\vspace{-5mm}
\caption{MacroPARAFAC residual maps of several samples with outlying behavior.}
\label{fig:Dorrit_ResidualMap_indiv}
\end{figure}

\begin{figure}[!ht]
\begin{center}
\includegraphics{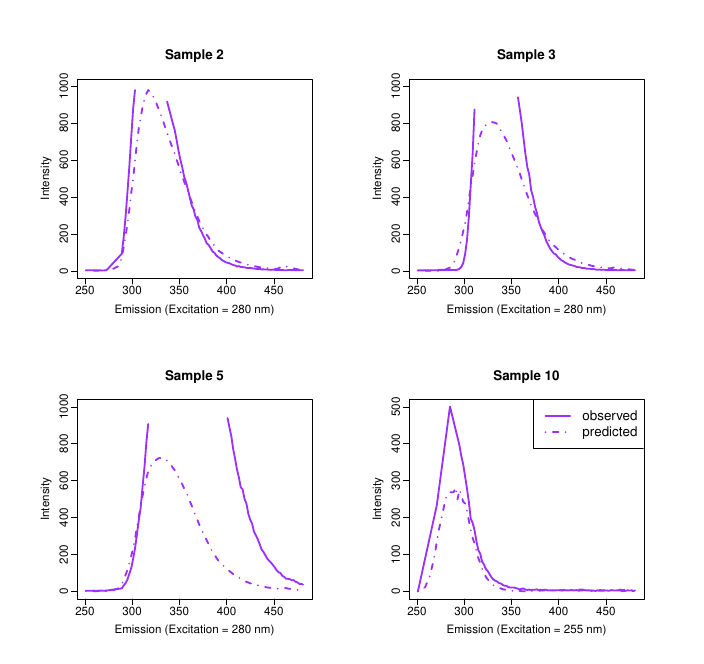} 
\end{center}
\vspace{-5mm}
\caption{Observed (full line) and predicted (dash-dotted line) MacroPARAFAC emission spectra of samples 2, 3 and 5 at excitation wavelength 280 nm, and of sample 10 at excitation wavelength 255 nm. The spectra of samples 2, 3 and 5 are interrupted as they are not fully observed.}
\label{fig:Dorrit_emission_samples}
\end{figure}

We study this in more detail by making 
residual maps of the samples 2, 3, 4, 5, 6, 10, 12 and 16. We thus focus on several horizontal slices of our data array and combine the 116 emission wavelengths into 29 blocks. Figure~\ref{fig:Dorrit_ResidualMap_indiv} displays the resulting residual maps. We immediately spot the missing values in samples 2, 3 and 5. Their large number of outlying residuals and their high degree of outlyingness is also very apparent. 
When we plot their observed and predicted emission spectrum at excitation wavelength 280 nm (Figure~\ref{fig:Dorrit_emission_samples}) 
we observe that the estimated intensity of their unobserved peak is considerably lower than what we would obtain through extrapolating the observed peak.

Samples $6$-$10$ expose large positive residuals at the lower excitation and emission wavelengths. It turns out that they have much larger concentrations of phenylalanine than the other samples, hence MacroPARAFAC considers the corresponding cells as outlying and yields lower estimates, as can be seen on the lower right panel of Figure~\ref{fig:Dorrit_emission_samples}. Consequently the estimated concentrations of the imputed data become regular, and so is their score distance.

\begin{figure}[!ht]
\begin{center}
    \includegraphics[scale=0.4]{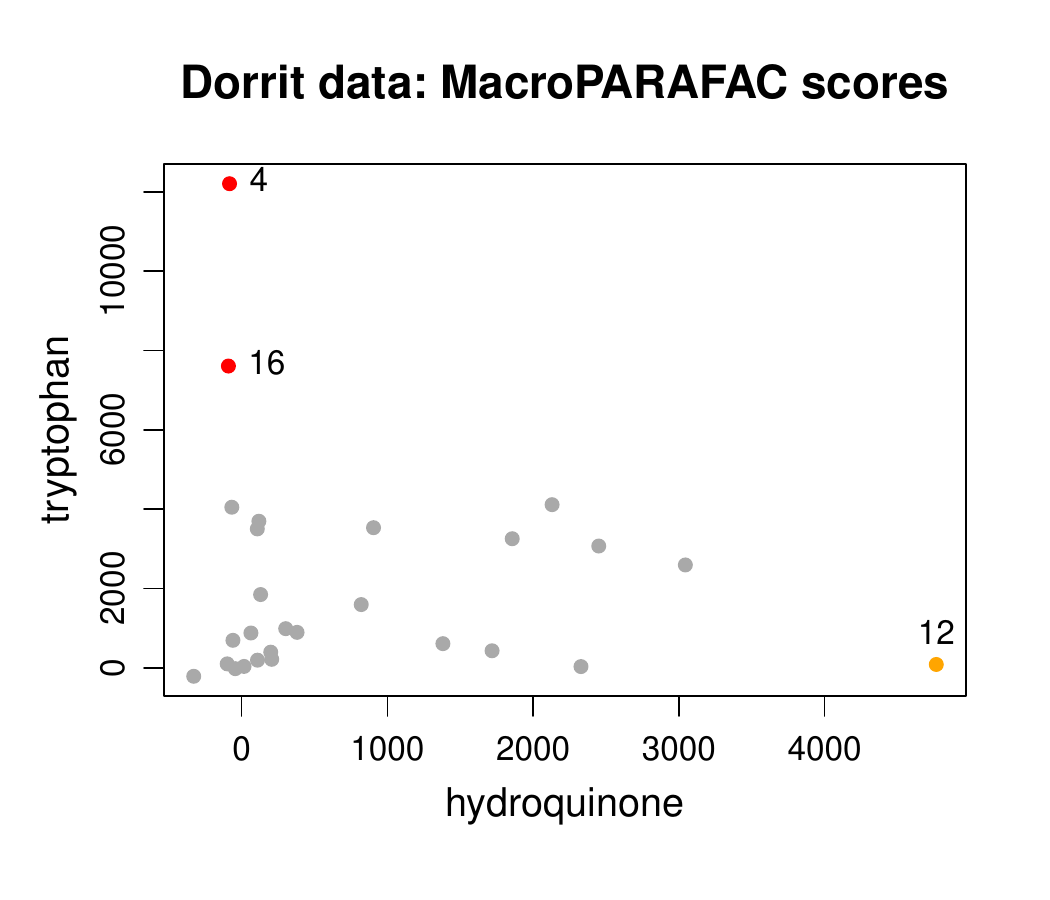}
\end{center}
\vspace{-10mm}
\caption{MacroPARAFAC scores plot of hydroquinone and tryptophan.}
\label{fig:Dorrit_MacroPAR_scores}
\end{figure}

Samples 4, 12 and 16 also exhibit an unusual large peak due to a large concentration of hydroquinone (12) and tryptophan (4, 16). As their amount of outlying cells is among the largest in the data set, MacroPARAFAC excludes them to estimate the loadings. 
In the final stage it turns out that nevertheless they fit the model quite well, but that they have large estimated concentrations, clearly seen on the plot of the MacroPARAFAC scores corresponding to these two fluorophores (Figure~\ref{fig:Dorrit_MacroPAR_scores}).   

To further explore the robustness of MacroPARAFAC towards cellwise outliers, we have reintroduced the Raman and Rayleigh scatter in the samples 4, 10, 18, 19 and 20.  
As can be seen in Supplementary Figure~\ref{fig:Dorrit_loadings_scatter} the resulting loadings of MacroPARAFAC are still close to the pure spectra.
On the resulting outlier map (Supplementary  Figure~\ref{fig:Dorrit_OutlierMap_scatter}) three more orange samples (observations 18, 19 and 20) can be spotted having a large RD but small imputed RD. Also the RD of samples 4 and 10 has increased, as expected. 

Also from the residual distance plot in Supplementary Figure~\ref{fig:Dorrit_ResidualDistanceMap_scatter} it can be derived that the scatter in these 5 samples is considered as cellwise contamination that can be removed by properly imputing the scatter. On the residual maps of the samples with scattering, displayed in Supplementary  Figure~\ref{fig:Dorrit_ResidualMap_indiv_scatter}, the scatter can be clearly seen as outlying cells.  

Following a reviewer's remark, we also illustrate the behavior of MacroPARAFAC at a skewed error distribution. Data were generated as $\ubX = \ubX_{\text{pure}} + \ubE$ where $\bX_{\text{pure}} = \bA_{\text{pure}} (\boldsymbol{C}_{\text{pure}} \odot \boldsymbol{B}_{\text{pure}})^{\prime}$ contains the pure concentrations and loadings of the Dorrit data. The error distribution $\bE$ is first taken as the Gaussian distribution with zero mean and variance equal to 23.95 (which is roughly the variance of the Dorrit residuals). Next we set $\bE$ equal to the mean-centered lognormal distribution $LN(\mu=0, \sigma^2=1.3)$ which has the same variance. Finally we considered an even more challenging setting where we increased the variance of the skewed errors by a factor of 25. Supplementary Figure~\ref{fig:Dorrit_Skew_Simulation} shows one sample for each of these settings, as well as the estimated emission loadings. We notice that the estimates become more wiggling when the errors are skewed, but they still follow the general shape of the true loadings. It would be worth investigating this behavior in more detail, but a more profound simulation study is outside the scope of this paper. Obviously also the outlier detection rules should be adapted following the distribution of the errors.

\section{Conclusion}
MacroPARAFAC introduces a new robust approach to estimate the loadings and scores within a PARAFAC model. It offers estimates for missing values, is resistant towards outlying cases and provides imputed values for outlying cells. Residual maps and the enhanced outlier map serve as powerful diagnostic tools to visualize the degree and the type of outlyingness of the observations and individual cells in the data array. This, in turn, enhances the interpretability of the estimated loadings and scores, facilitating improved detection and analysis of unusual behavior. 
The R code of MacroPARAFAC and several graphical tools will become available as part of the R package \texttt{cellWise}.

We envision extending our methodology to encompass other dimension reduction models for multiway data, including the Tucker3 model \citep{Tucker:MathNotes}. 
This novel approach may also enhance the robust classification of multiway data.

\section*{Acknowledgement} We thank Wouter Saeys and Peter Rousseeuw for useful suggestions and feedback.

\bibliographystyle{Chicago}

\clearpage
\pagenumbering{arabic}
\appendix
\begin{center}
\large{Supplementary figures to: 
  MacroPARAFAC for handling rowwise\\ and cellwise outliers in incomplete multiway data}\\
\vspace{7mm}
\normalsize{Mia Hubert, 
  Mehdi Hirari} 
\end{center}
\vspace{3mm}

\renewcommand{\thesection}{S.\arabic{section}}
\renewcommand\thefigure{S.\arabic{figure}}   
\numberwithin{equation}{section} 
\renewcommand{\theequation}{S.\arabic{equation}} 
 
\setcounter{figure}{0}    
\setcounter{equation}{0}    

\section{Additional figures of the simulation study}

\begin{figure}[!htb]   
\centerline{%
\includegraphics[height=.25\paperheight]{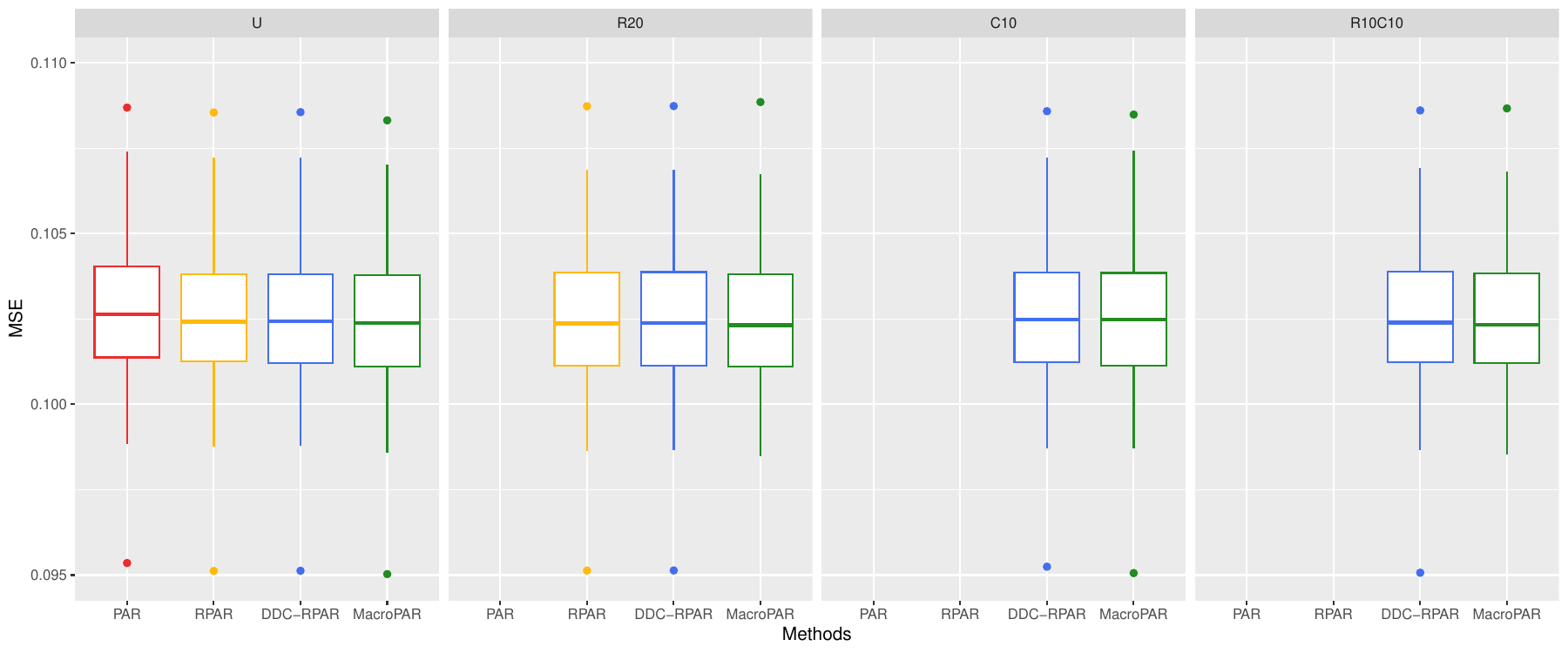}%
}%
\caption{Boxplots of MSE for simulated data under varying contamination schemes without missing values, truncated at MSE $ = 0.11$.}
\label{fig:simul_boxplotMSE_noMISS_trunc}
\end{figure}

\begin{figure}[!htb]   
\centerline{%
\includegraphics[height=.25\paperheight]{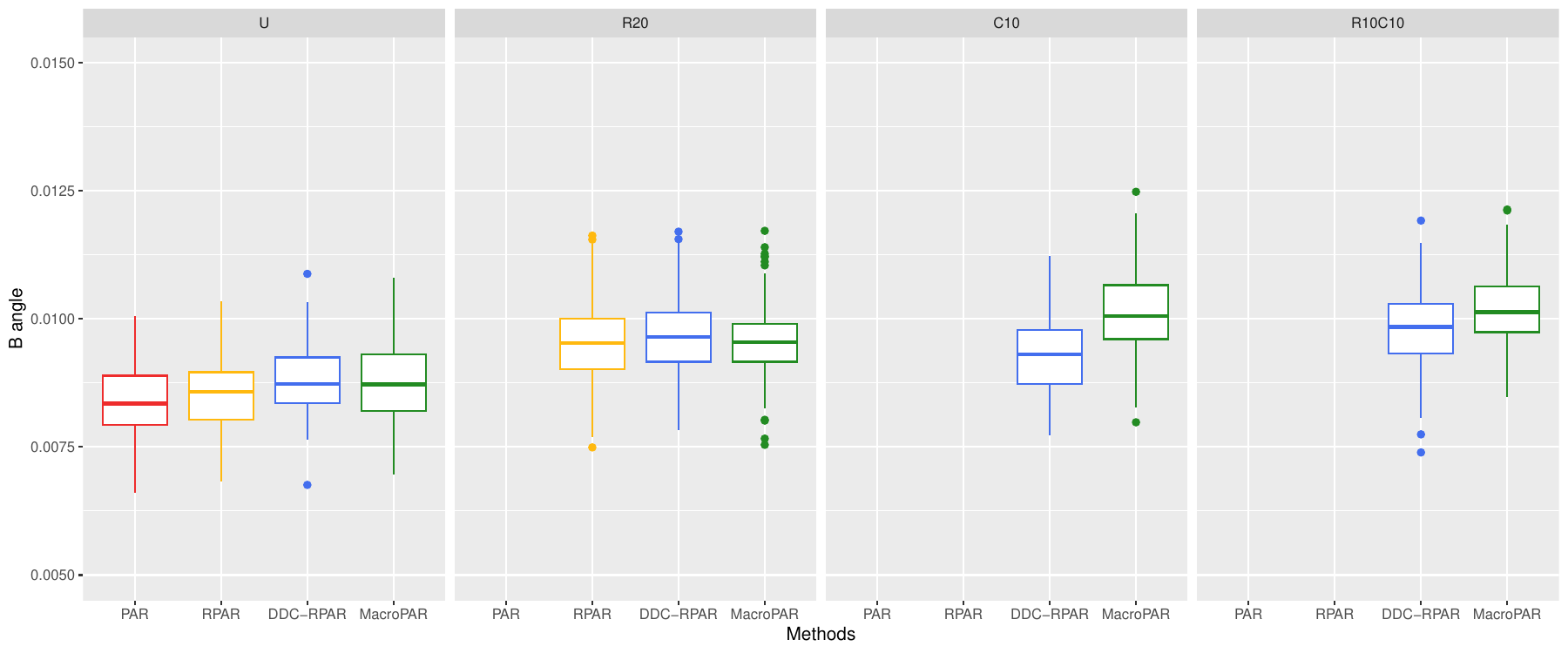}%
}%
\caption{Boxplots of $\bB$-angle for simulated data under varying contamination schemes without missing values, truncated at $0.015$.}
\label{fig:simul_boxplotBangles_noMISS_trunc}
\end{figure}

\begin{figure}[!htb]   
\centerline{%
\includegraphics[height=.25\paperheight]{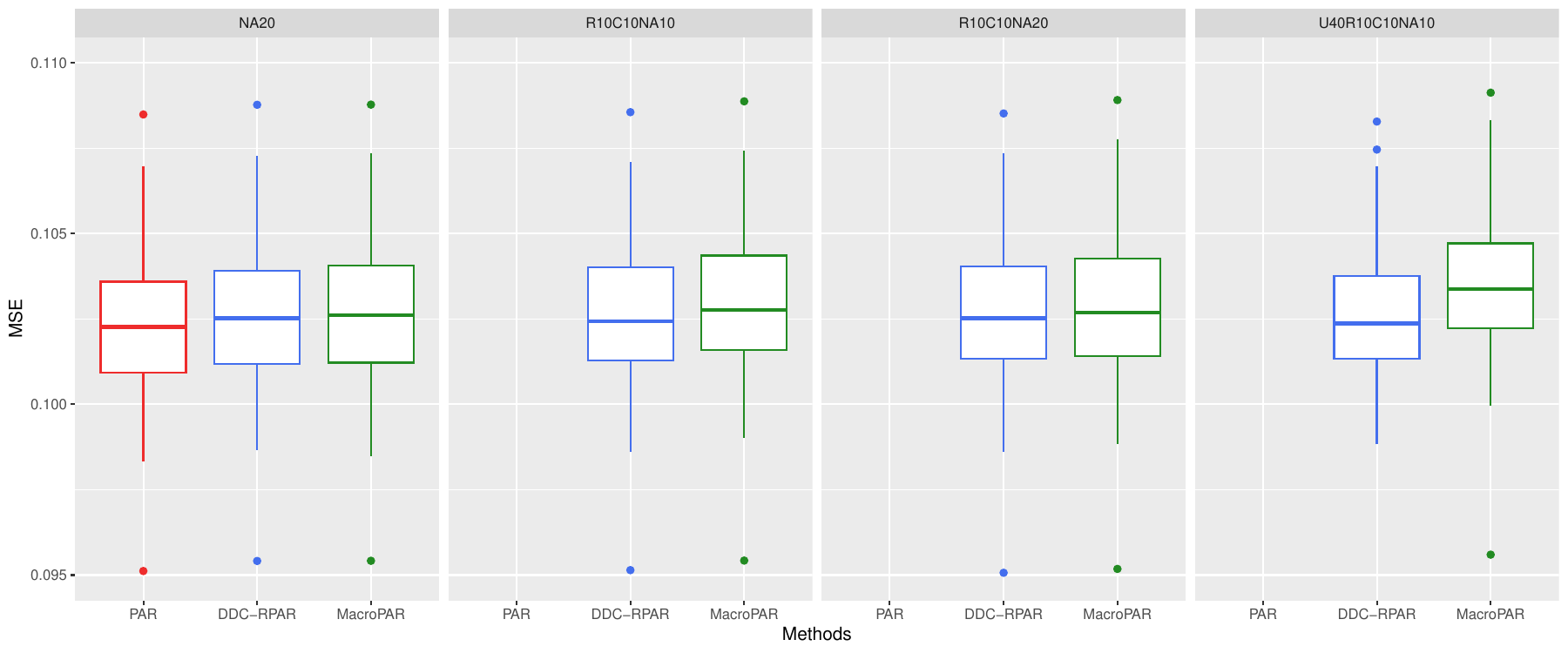}%
}%
\caption{Boxplots of MSE for simulated data under varying contamination schemes with missing values, truncated at MSE $ = 0.11$.}
\label{fig:simul_boxplotMSE_MISS_trunc}
\end{figure}

\begin{figure}[!htb]   
\centerline{%
\includegraphics[height=.25\paperheight]{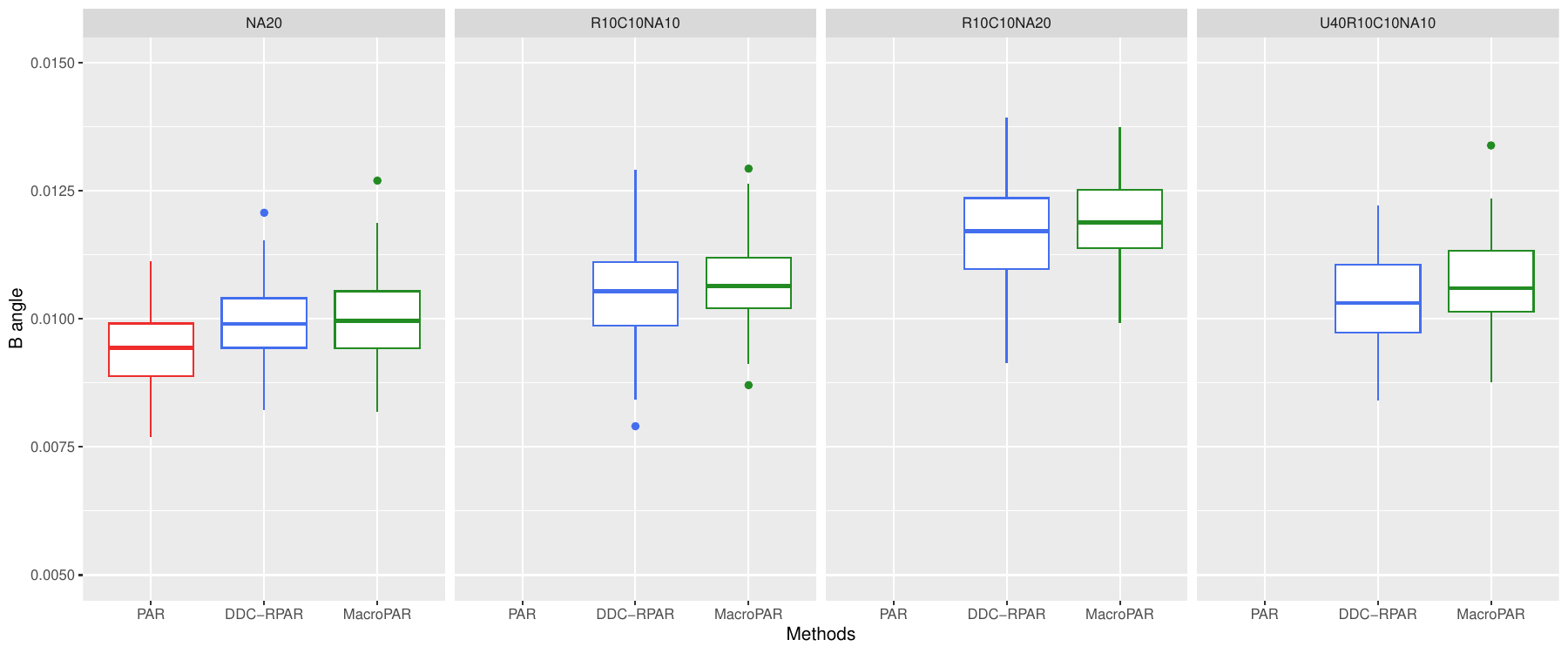}%
}%
\caption{Boxplots of $\bB$-angle for simulated data under varying contamination schemes with missing values, truncated at $0.015$.}
\label{fig:simul_boxplotBangles_MISS_trunc}
\end{figure}

\begin{figure}[!ht]
\begin{center}
\begin{tabular}{cc}
    \includegraphics[scale=0.4, trim=0cm 0cm 0cm 0cm, clip]{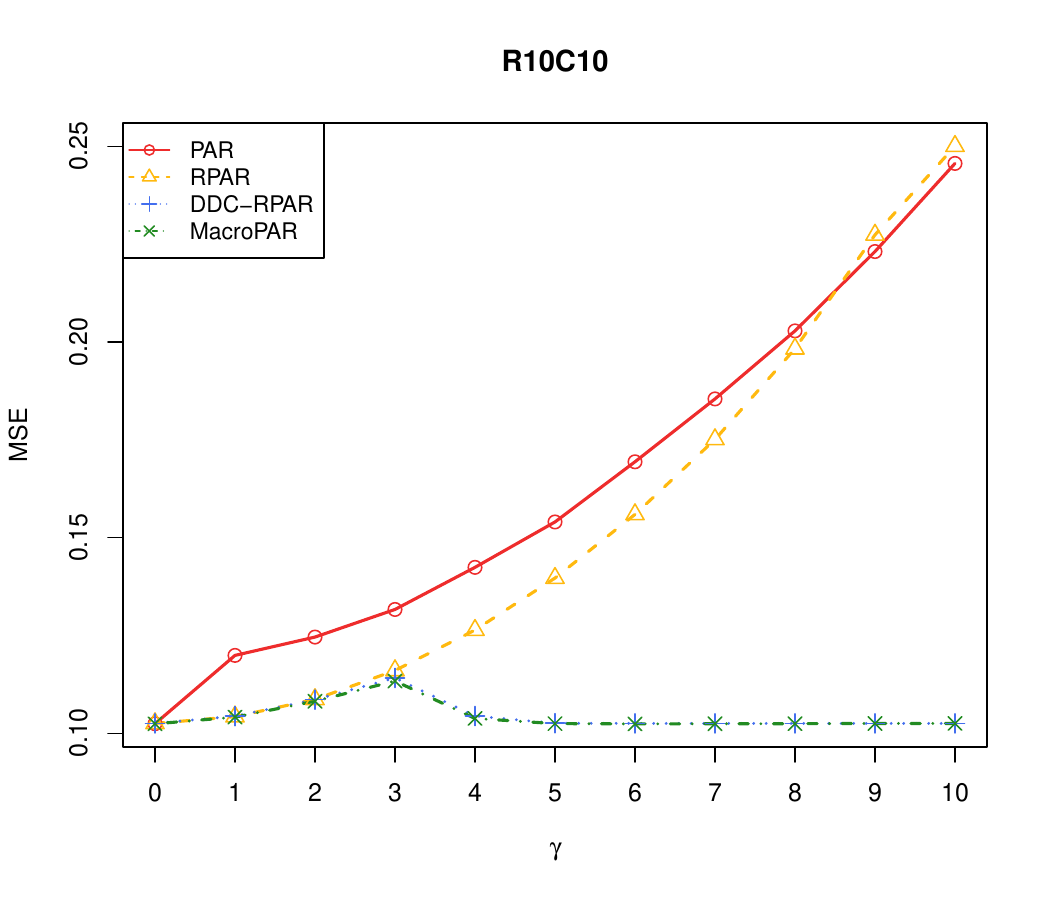} &
    \includegraphics[scale=0.4, trim=0cm 0cm 0cm 0cm, clip]{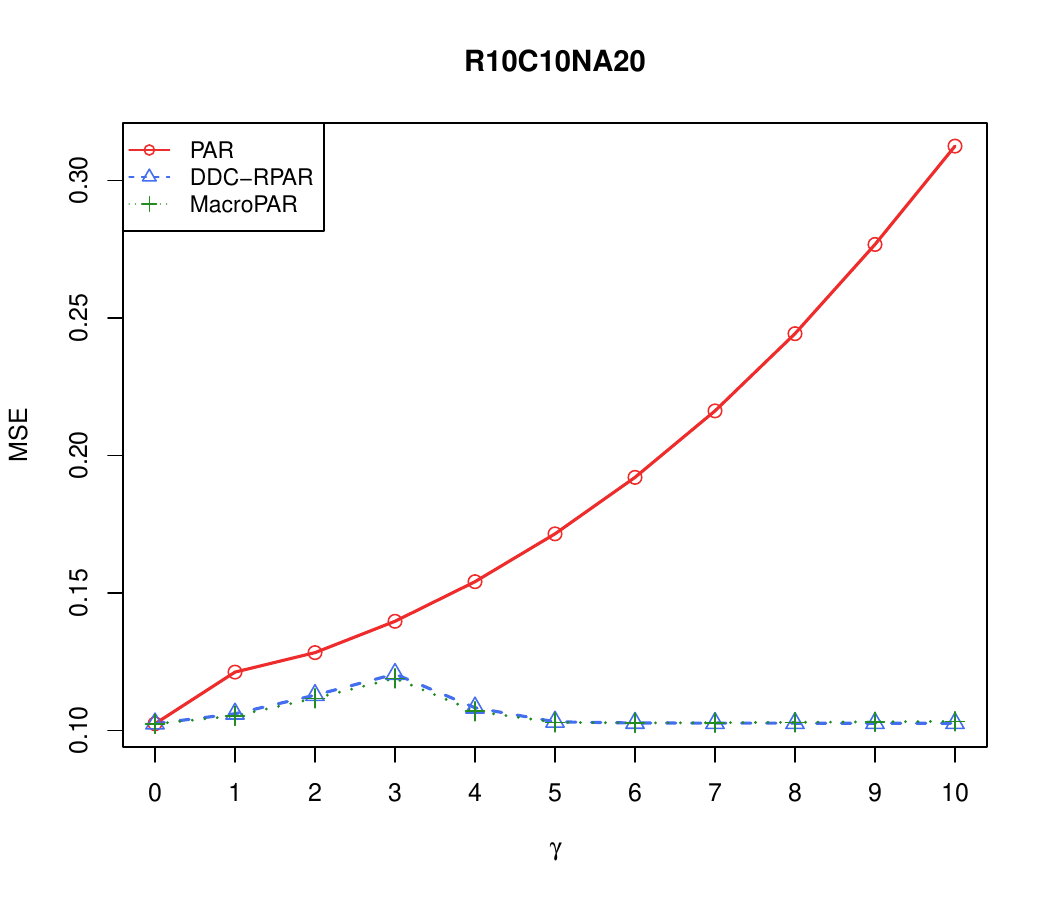} 
    \end{tabular}
\end{center}
\vspace{-10mm}
\caption{Average MSE curves for $\gamma$ ranging from 0 to 10.}
\label{fig:Simulation_MSEevol}
\end{figure}

\begin{figure}[!ht]
\begin{center}
    \includegraphics[scale=0.55]{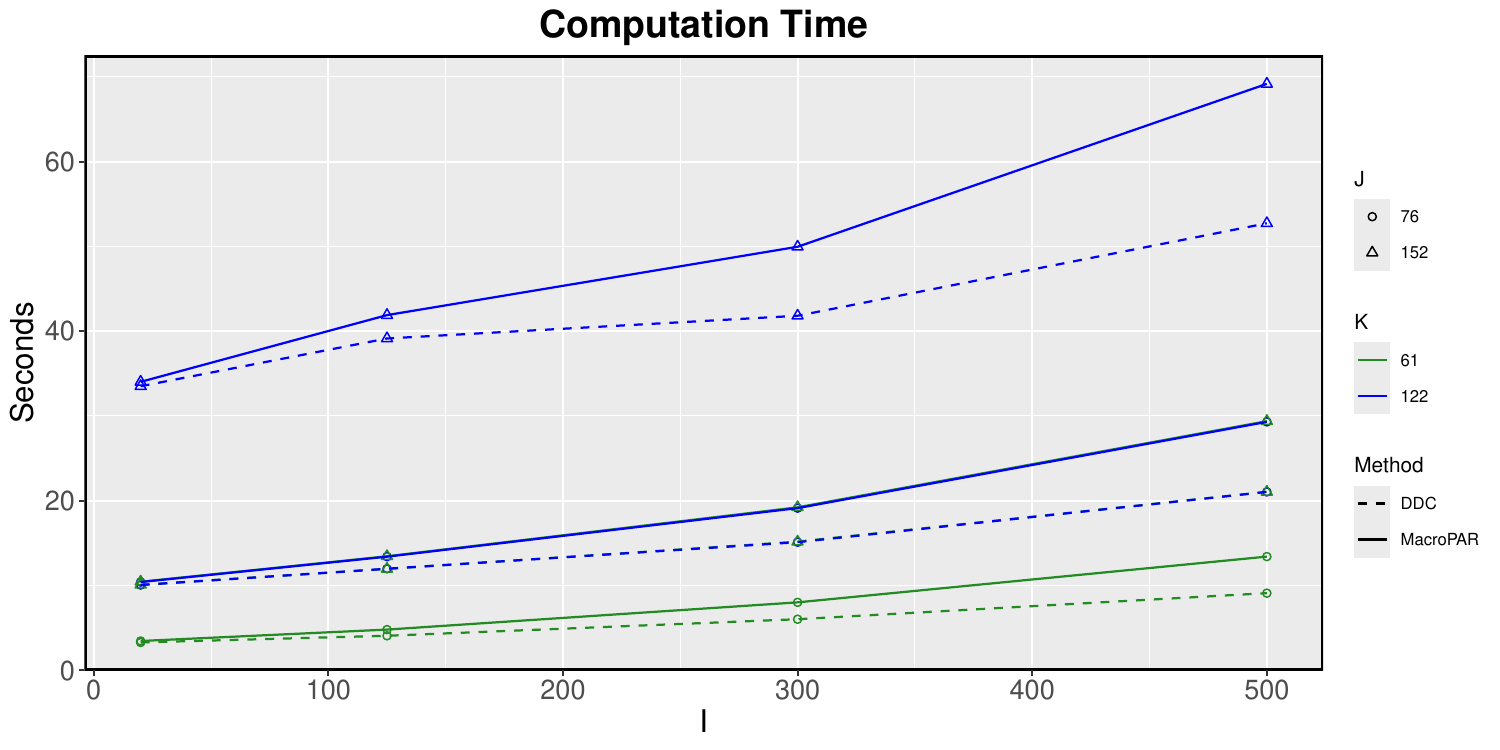}
\end{center}
\vspace{-3mm}
\caption{DDC and MacroPARAFAC computation times for varying values of $I, J$ and $K$.}
\label{fig:Simulation_MacroPAR_CompTime}
\end{figure}

\clearpage
\section{Additional figures of the Dorrit data analysis}

\begin{figure}[!ht]
\begin{tabular}{cc}
\includegraphics[width=7cm]{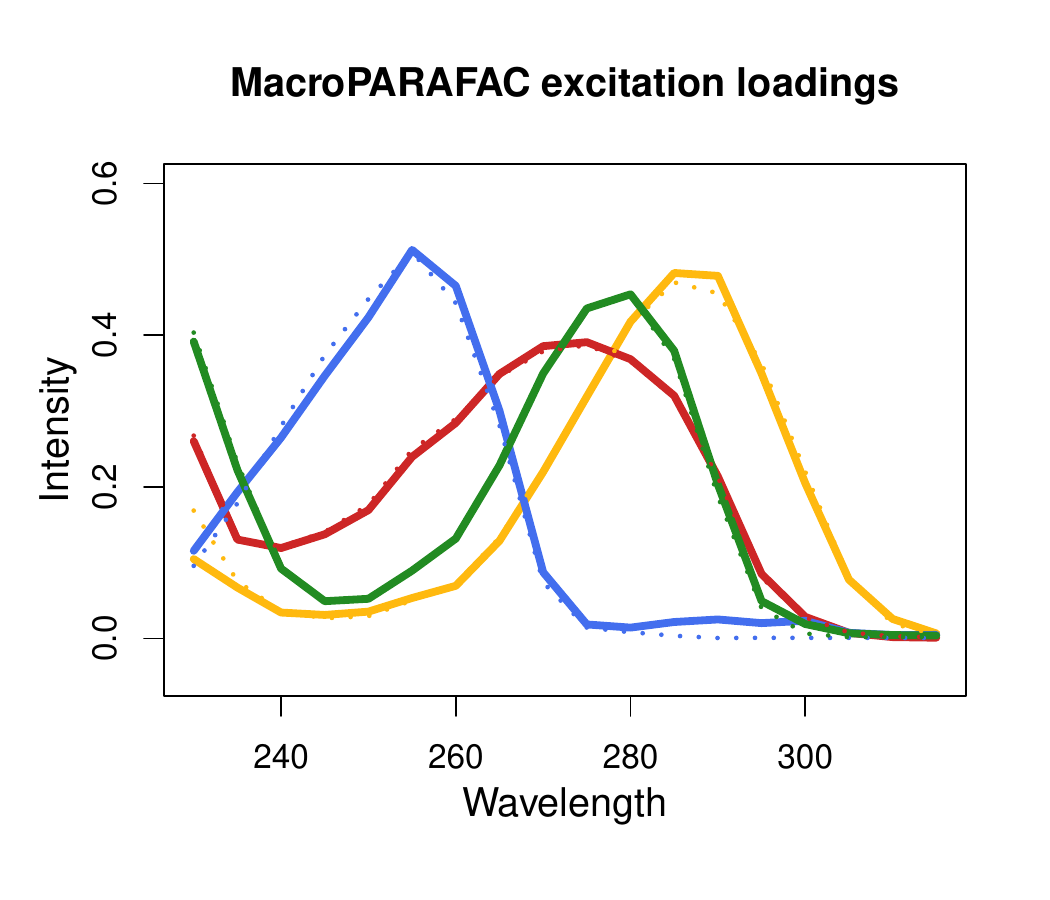} &
\includegraphics[width=7cm]{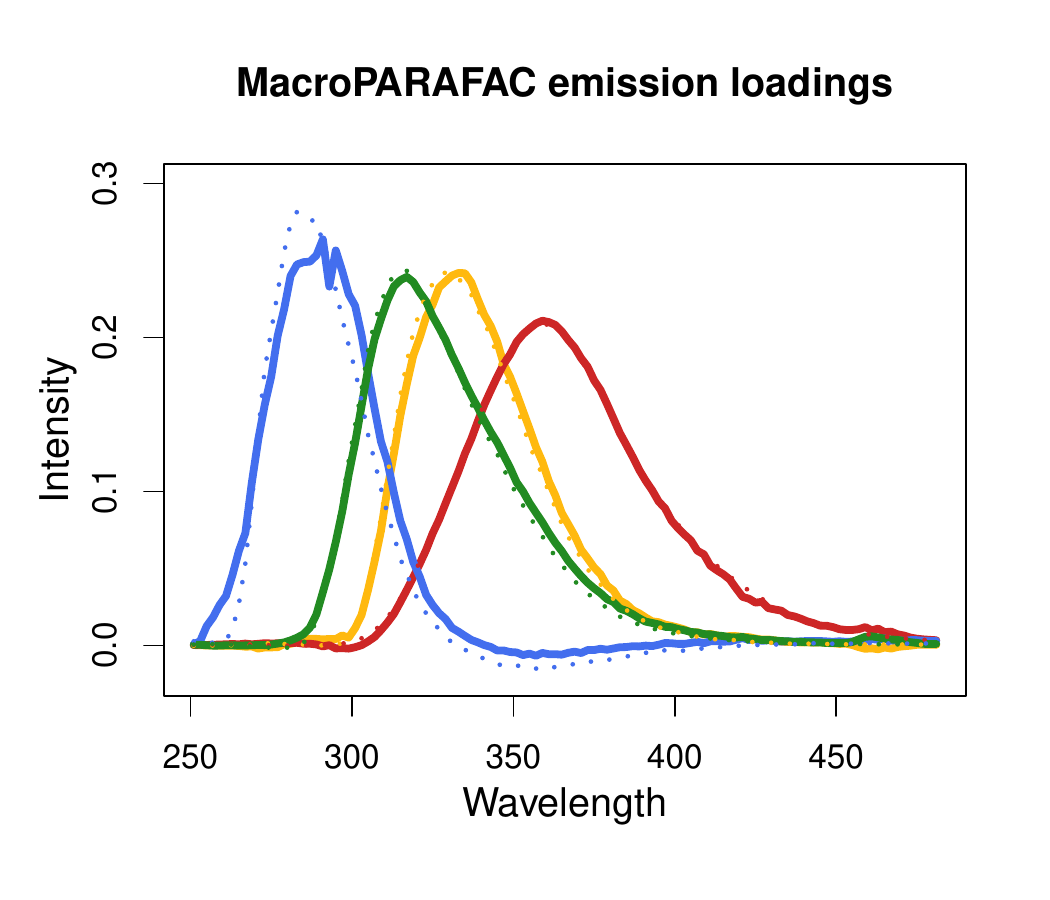} 
\end{tabular}
\vspace{-3mm}
\caption{MacroPARAFAC excitation  (left) and emission (right) spectra from the Dorrit data with scattering. }
\label{fig:Dorrit_loadings_scatter}
\end{figure}

\vspace{2cm}

\begin{figure}[!ht]
\begin{center}
    \includegraphics[scale=0.45]{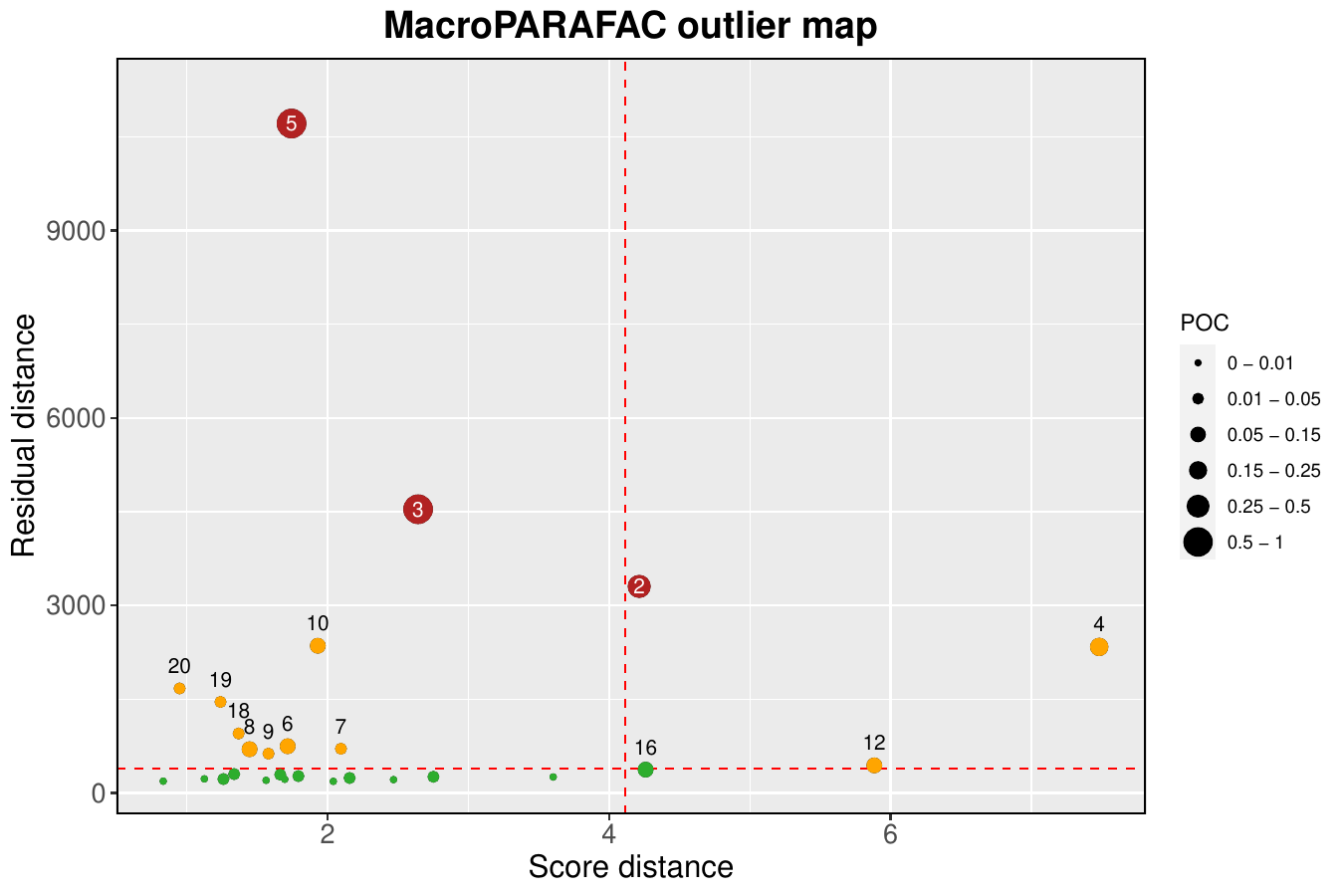}
\end{center}
\vspace{-3mm}
\caption{Enhanced outlier map of the Dorrit data with scattering in 5 samples.}
\label{fig:Dorrit_OutlierMap_scatter}
\end{figure}

\newpage
\begin{figure}[!ht]
\begin{center}
    \includegraphics[scale=0.5]{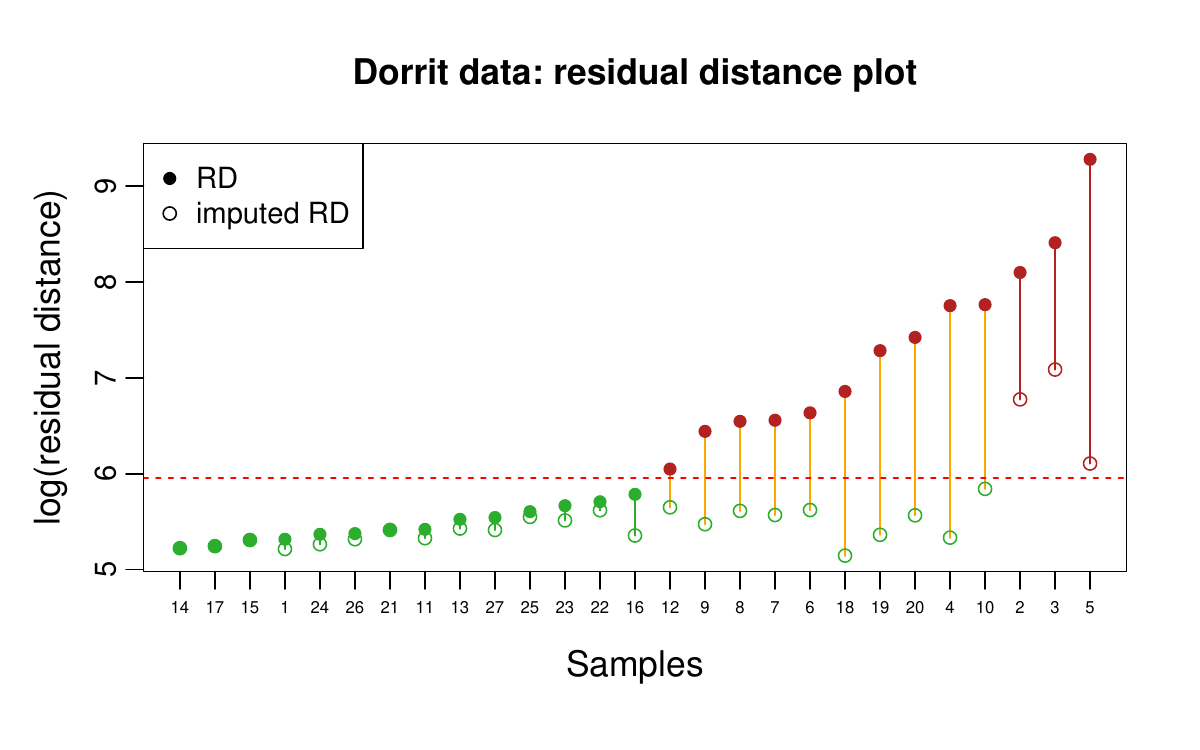}
\end{center}
\vspace{-7mm}
\caption{Plot of the MacroPARAFAC residual distances and imputed residual distances of the Dorrit data with scattering.}
\label{fig:Dorrit_ResidualDistanceMap_scatter}
\end{figure}

\vspace{1cm} 

\begin{figure}[!ht]
\begin{center}
    \includegraphics[scale=0.6, trim=0cm 0cm 0cm 1cm, clip]{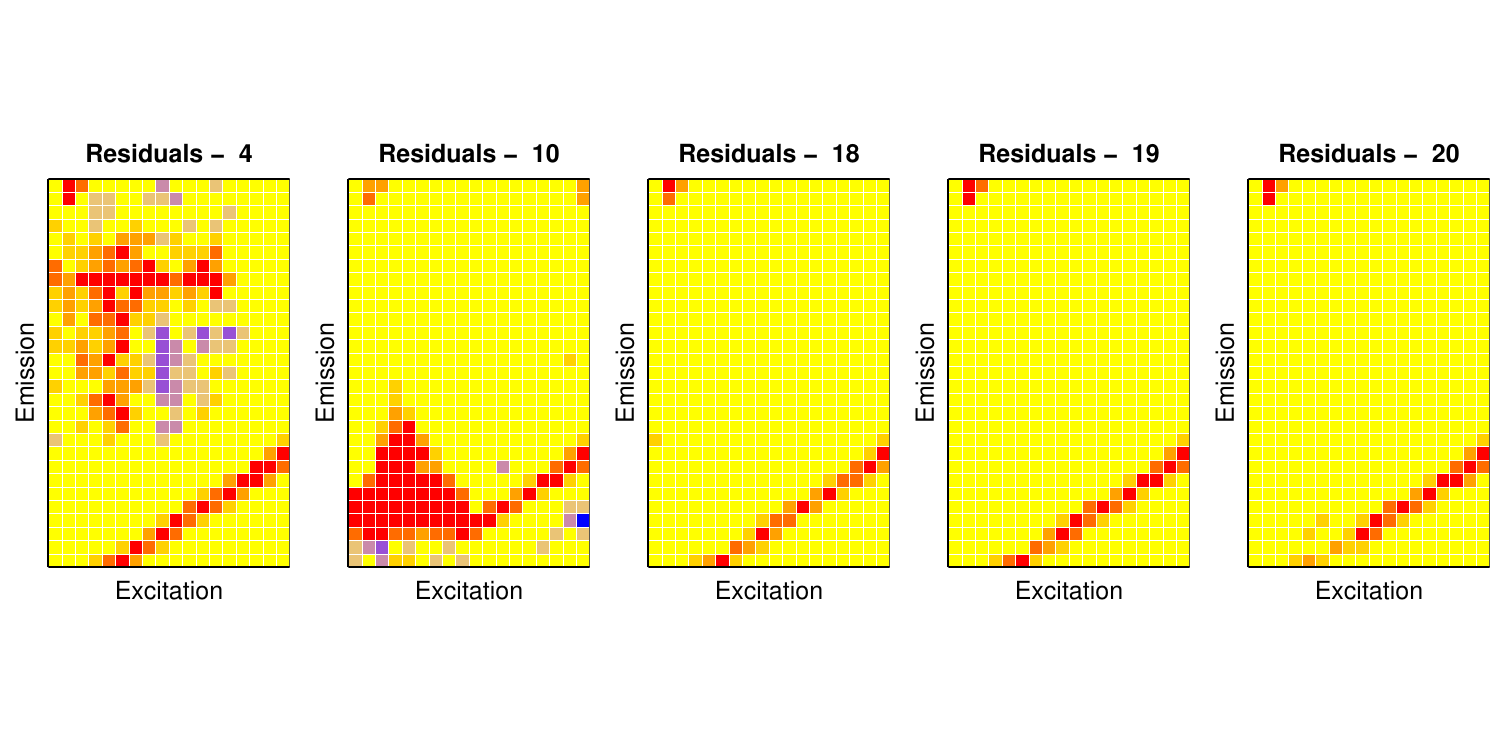}
\end{center}
\vspace{-15mm}
\caption{MacroPARAFAC residual maps of the samples with scattering.}
\label{fig:Dorrit_ResidualMap_indiv_scatter}
\end{figure}

\begin{figure}[!ht]
\centering
\begin{tabular}{cc}
\hspace{-1cm}\includegraphics[scale=0.6]
{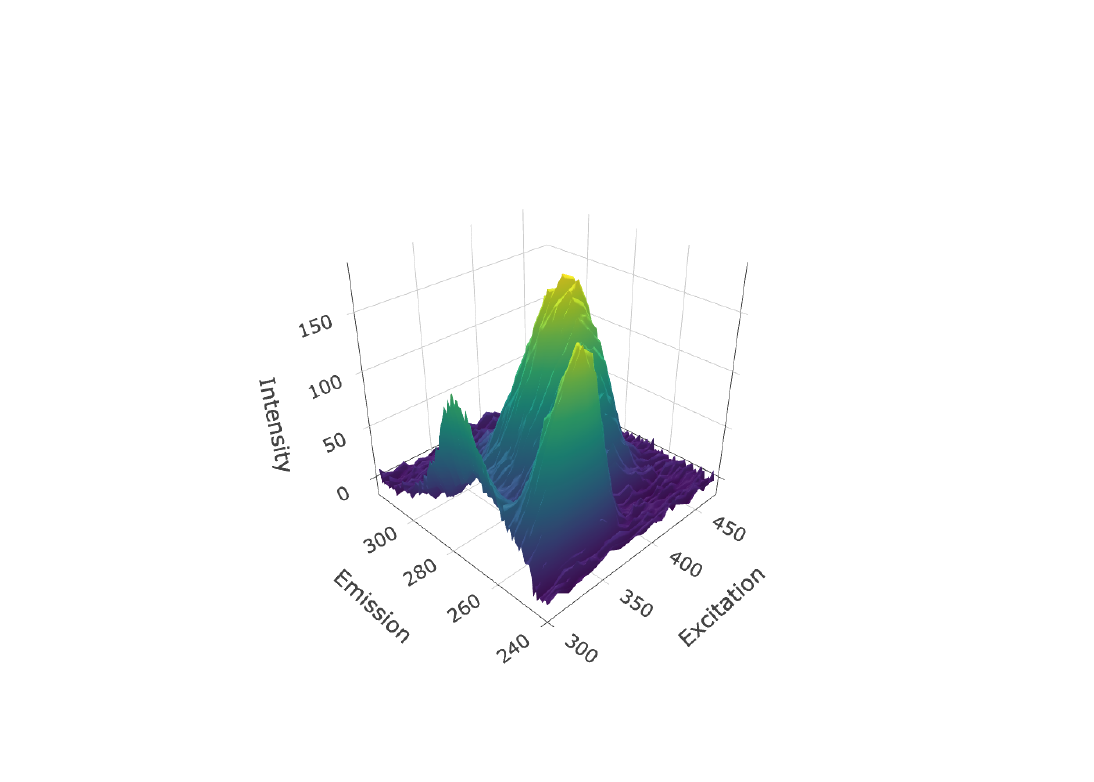} &
\hspace{-2cm}\includegraphics[scale=0.4]{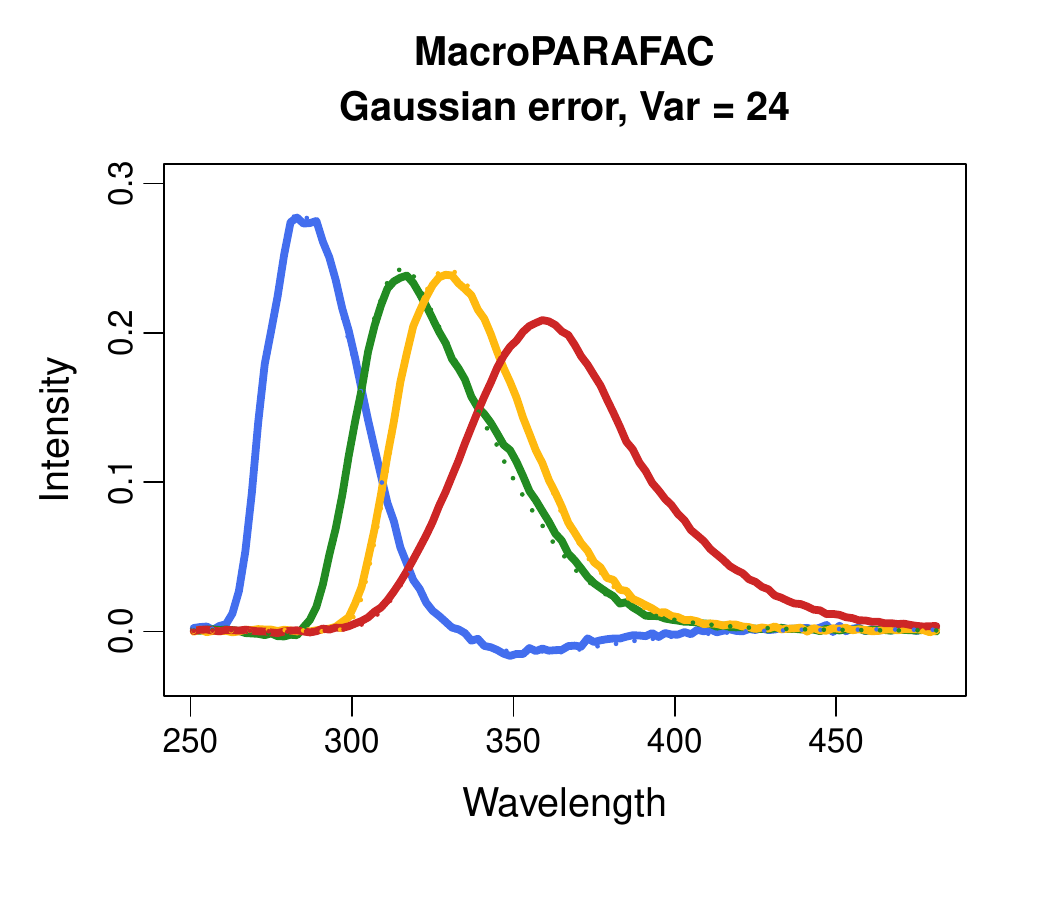} \\[-2cm]
\hspace{-1cm}\includegraphics[scale=0.6]
{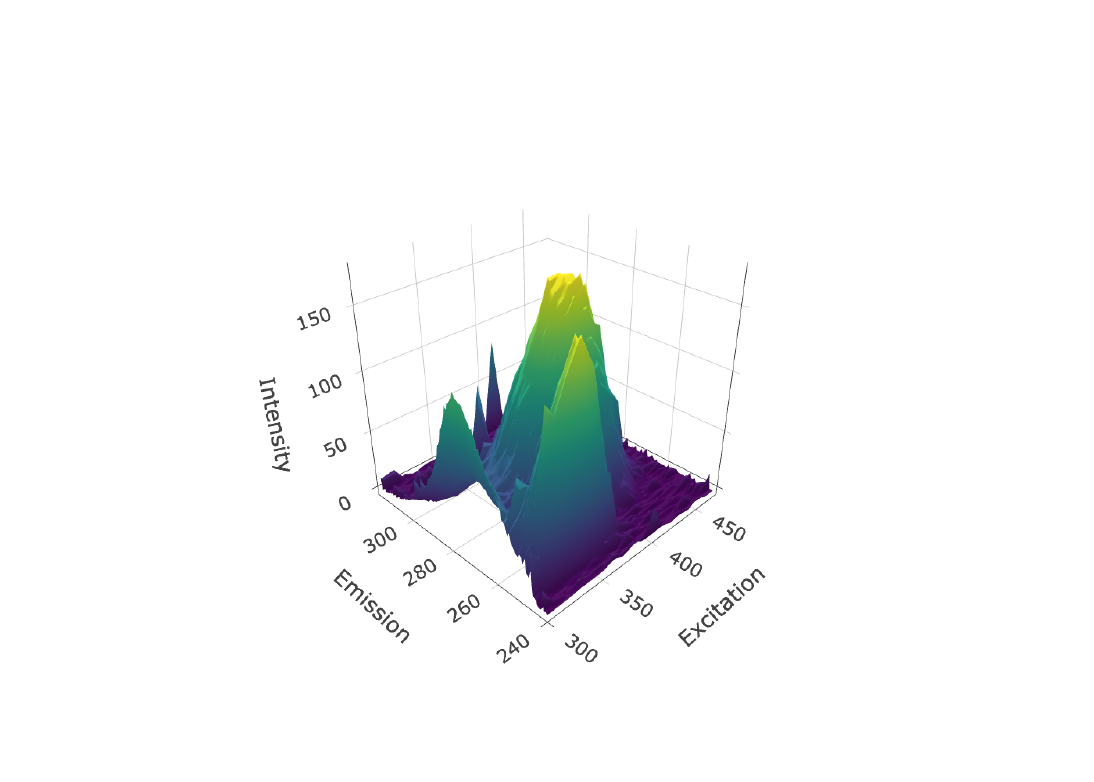} &
\hspace{-2cm}\includegraphics[scale=0.4]{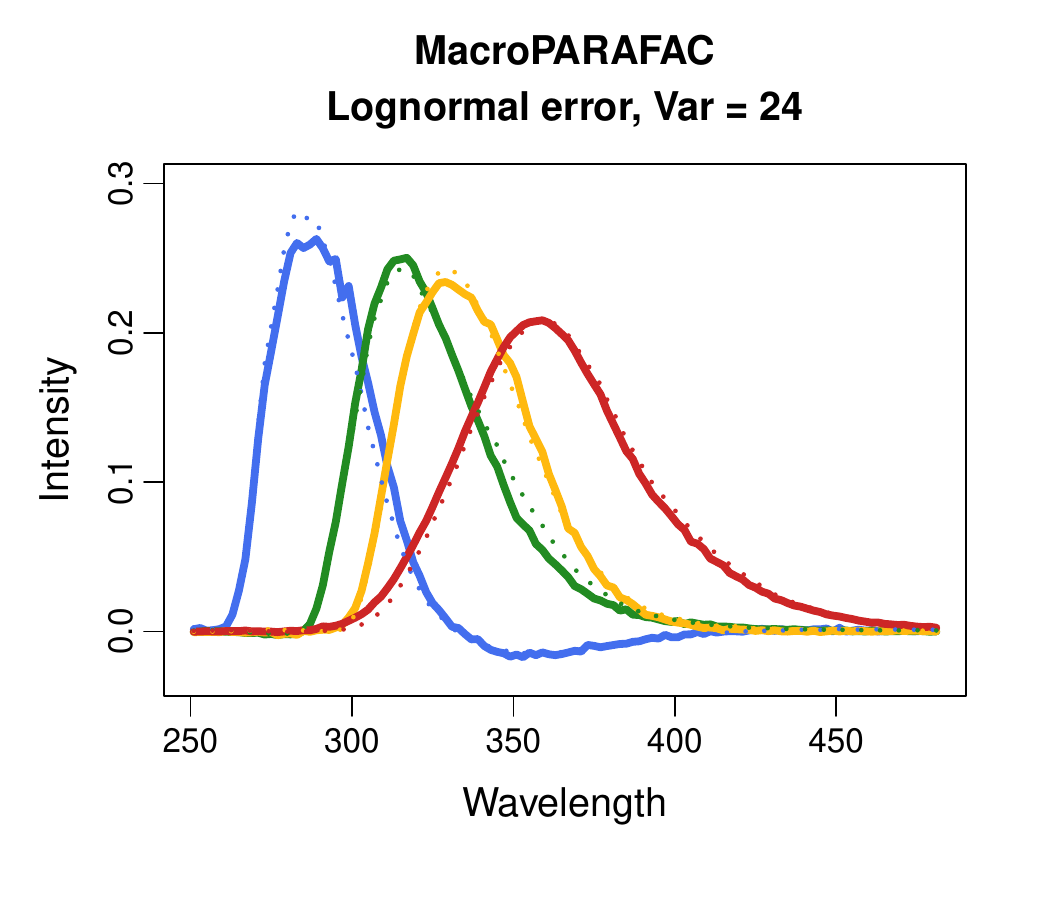} \\
[-2cm]
\hspace{-1cm}\includegraphics[scale=0.6]{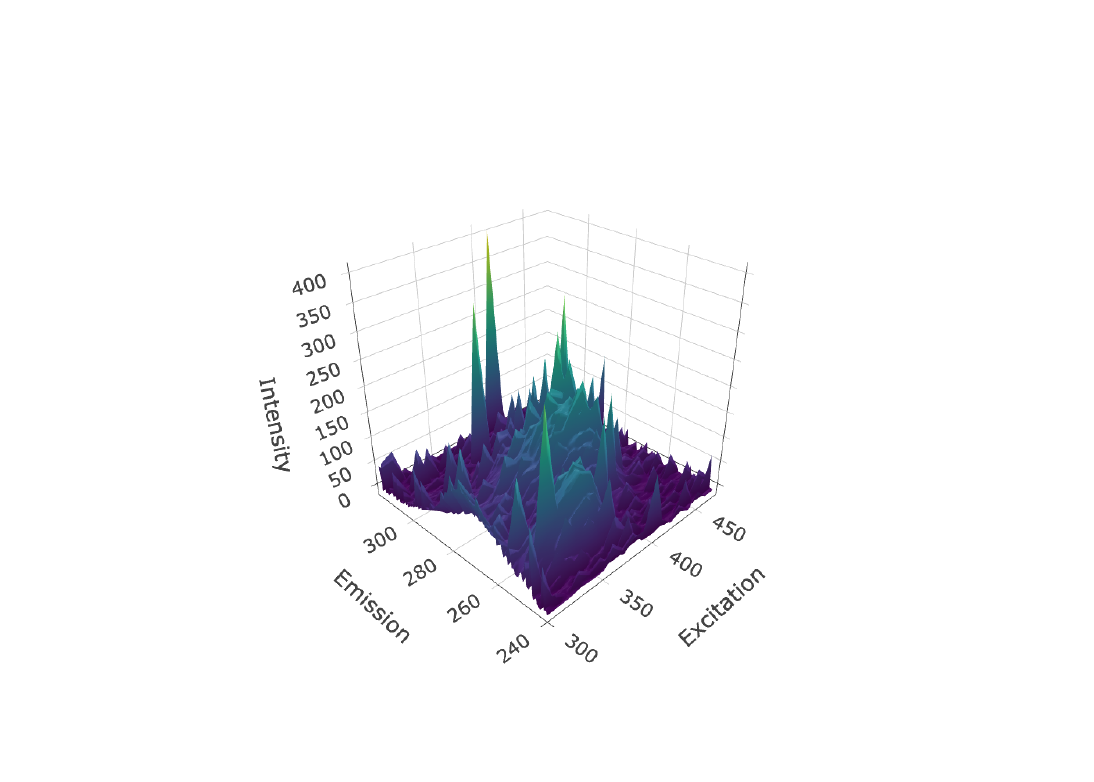} &
\hspace{-2cm}\includegraphics[scale=0.4]{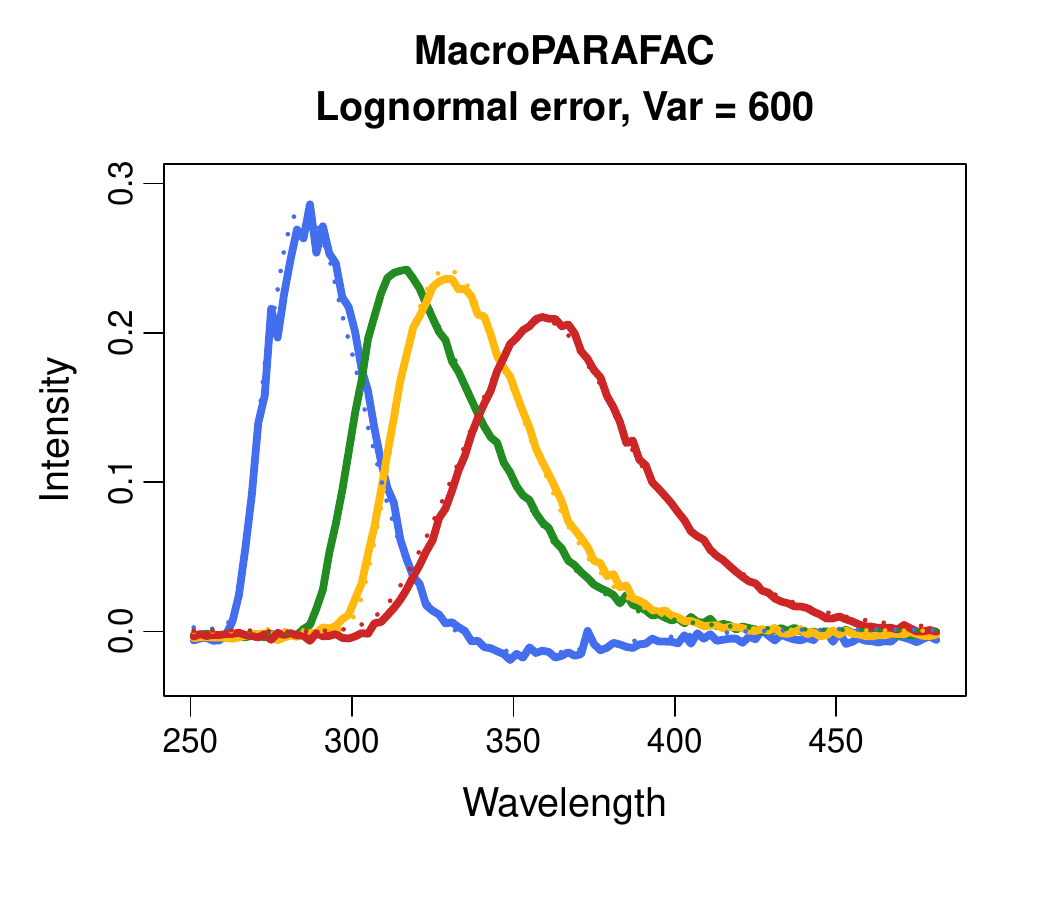} \\
\end{tabular}
\vspace{-3mm}
\caption{Landscapes (left) and MacroPARAFAC emission loadings (right) using Gaussian errors (top), lognormal errors with the same variance (middle) and lognormal errors with very large variance (bottom).}
\label{fig:Dorrit_Skew_Simulation}
\end{figure}

\end{document}